\documentclass[aps, prd, preprintnumbers, showpacs, nofootinbib, amsmath, amssymb, 
floatfix,
onecolumn, letterpaper, 
noshowpacs, showkeys, superscriptaddress]{revtex4}
\usepackage{amsfonts}
\usepackage{amsmath}
\usepackage{amssymb,epsf}
\usepackage{latexsym}
\usepackage{graphicx,epsfig}
\usepackage{amssymb}
\usepackage{subfigure}
\usepackage[citecolor=blue,linkcolor=black,urlcolor=blue,colorlinks=true,filecolor=green,
pdfstartview=FitV, 
pdftitle={},
pdfauthor={Mojtaba Najafizadeh},
pdfsubject={},
pdfkeywords={},
pdfpagemode=None,
bookmarksopen=true]{hyperref}
\usepackage{epstopdf}
\usepackage{color}
\usepackage[normalem]{ulem}
\usepackage{graphicx}
\usepackage{graphicx}
\graphicspath{ {images/} }
\usepackage{commath}
\usepackage{amsthm,amssymb,amsmath,mathrsfs}
\usepackage{mathtools}
\usepackage{empheq}
\usepackage{tabu} 
\usepackage[most]{tcolorbox}

\DeclarePairedDelimiterX\braket[2]{\langle}{\rangle}{#1 \delimsize\vert #2}
\graphicspath{ {images/} }

\def\be {\begin{equation}}
\def\ee {\end{equation}}
\def\bea {\begin{eqnarray}}
\def\eea {\end{eqnarray}}
\def\bc {\begin{center}}
	\def\ec {\end{center}}
\def\bg {\begin{align}}
\def\eg {\end{align}}
\def\bi {\begin{itemize}}
	\def\ei {\end{itemize}}

\def\le {\left}
\def\ri {\right}
\def\p {\partial}

\def\vs {\vspace}

\def\ds{\partial\hspace{-6pt}\slash\hspace{0pt}}

\def\ds{\partial\hspace{-6pt}\slash\hspace{1pt}}
\def\ws{\omega\hspace{-7pt}\slash\hspace{3pt}}
\def\dws{\partial_\omega\hspace{-11pt}\slash\hspace{7pt}}
\def\es{\eta\hspace{-5pt}\slash\hspace{0pt}}

\def\c  {\cdot}
\def\a  {\alpha}
\def\b  {\beta}
\def\g  {\gamma}
\def\d  {\delta}

\def\e  {\eta}
\def\de {\partial_\eta}
\def\m  {\mu}
\def\n  {\nu}
\def\r  {\rho}
\def\w {\omega}
\def\dw{\p_\omega}
\def\dww{\p_\omega^{\,2}}

\def\s {\sigma}
\def\wdx {(\omega\cdot\p_x)}
\def\dwdx {(\dw\cdot\p_x)}

\begin{document}                             

\title{\LARGE{ Supersymmetric Continuous Spin Gauge Theory
\vskip .2cm}} 




\author{Mojtaba Najafizadeh}
\email{mnajafizadeh@ipm.ir}
\affiliation{School of Physics, Institute for Research in Fundamental Sciences (IPM), \\ P.O.Box 19395-5531, Tehran, Iran}

\begin{abstract}
Taking into account the Schuster-Toro action and its fermionic analogue discovered by us, we supersymmetrize unconstrained formulation of the continuous spin gauge field theory. Afterwards, building on the Metsaev actions, we supersymmetrize constrained formulation of the theory. In each formulation, we provide supersymmetry transformations for the $\mathcal{N}=1$ supermultiplet in four-dimensional flat space-time, in which continuous spin particle (CSP) is considered to be a complex scalar continuous spin field, and its superpartner which can be called ``\,CSPino\,'' is considered to be a Dirac continuous spin field. It is shown that the algebra of these supersymmetry transformations are closed on-shell. Furthermore, we investigate whether obtained supersymmetry transformations reproduce the known result of the higher spin gauge field theory in the helicity limit. Finally, we illustrate how these two separate set of obtained transformations are related to each other. 
\end{abstract}
%
\preprint{IPM/P-2019/046}
\vspace{-1.1cm}
\maketitle
\rule{\textwidth}{.4pt}
\vspace{-1.1cm}
{\small\tableofcontents}

%
\section{Introduction}
Elementary particles propagating on Minkowski space-time have been classified long time ago by Wigner using the unitary irreducible representations (UIRs) of the Poincar\'e group $ISO(3,1)$\,\cite{Wigner}\,(see also \cite{Unitary} for more details in any dimension). In $d$ space-time dimensions, the massive particles are determined by representations of the rotation group $SO(d - 1)$, while the massless particles (helicity particles) which describe particles with a finite number of degrees of freedom are determined by representations of the Euclidean group $E_{d-2}=ISO(d - 2)$. Another massless representation, called continuous spin representation\,{\color{blue}\footnote{Also known as infinite spin representation in the literature.}}, describes a continuous spin particle (CSP) with an infinite number of physical degrees of freedom per spacetime point characterized by the representations of the short little group $SO(d - 3)$, the little group of $E_{d - 2}$\,\cite{Brink:2002zx}. This representation labels by a dimensionful parameter $\m$ (a real parameter
with the dimension of a mass) so as when $\m$ vanishes, the helicity eigenstates do not mix while they do when $\m\neq0$. Thus, the continuous spin parameter $\m$ controls the degree of mixing. In fact, in the ``helicity limit'' $\m\rightarrow0$, the continuous spin representation becomes reducible and decomposes into the direct sum of all helicity representations. We recall that for the both massless representations, helicity and continuous spin, the eigenvalue of the quadratic Casimir operator $C_2 := P^2 $ (the square of the momentum $P_\m$) vanishes. However, for the helicity representation, the eigenvalue of the quartic Casimir operator $C_4 := W^2$ (the square of the Pauli-Lubanski vector ${W}^\m= \frac{1}{2} \, \epsilon^{\m\n \rho\sigma} \, {P}_\n \, {J}_{\rho\sigma}$) is zero, while the one for the continuous spin representation becomes $\m^2$.

\vspace{.2cm}

Historically, constructing a local covariant action principle for continuous spin particle has been a mystery for decades, however, about 75 years after Wigner's classification, the first action principle for the bosonic continuous spin particle was presented by Schuster and Toro \cite{Schuster:2014hca} in 2014, and the first action principle for the fermionic continuous spin particle was suggested in 2015 \cite{Najafizadeh:2015uxa}. In these two action principles there are no constraint on the gauge fields and parameters, so in this sense one can refer to them as unconstrained formulations of the CSP theory. Along with the unconstrained formulation, Metsaev established a constrained formulation of the CSP theory for both the bosonic \cite{Metsaev:2016lhs} and fermionic \cite{Metsaev:2017ytk} continuous spin fields, in $d$-dimensional (A)dS space-time, in which the gauge fields and parameters are constrained. These two formulations of the CSP theory, unconstrained and constrained\,{\color{blue}\footnote{Notice that, in 4-dimensional flat space-time, we have explained in \cite{Bekaert:2017xin} how to obtain unconstrained formulation of the bosonic \cite{Schuster:2014hca} and fermionic \cite{Najafizadeh:2015uxa} CSPs, and moreover, we have elaborated in \cite{Najafizadeh:2017acd} how to acquire constrained formulation of the bosonic \cite{Metsaev:2016lhs} and fermionic \cite{Metsaev:2017ytk} CSPs, both directly from the Fronsdal-like and Fang-Fronsdal-like equations \cite{Bekaert:2005in}.}}$^,${\color{blue}\footnote{Note also that in the helicity limit $\m\rightarrow 0$, unconstrained formulation of the bosonic \cite{Schuster:2014hca} and fermionic \cite{Najafizadeh:2015uxa} CSP actions reproduce, respectively, the bosonic \cite{Segal:2001qq} and fermionic \cite{Najafizadeh:2018cpu} higher spin actions, and constrained formulation of the bosonic \cite{Metsaev:2016lhs} and fermionic \cite{Metsaev:2017ytk} CSP actions reproduce, separately, the Fronsdal \cite{Fronsdal:1978rb} and Fang-Fronsdal \cite{Fang:1978wz} actions, in 4-dimensional flat space-time.\label{foot}}}, that have been formulated based on the metric-like approach, have not yet been supersymmetrized in the literature, which is the main purpose of the present paper. Indeed, for each formulation, we provide supersymmetry transformations for the $\mathcal{N}=1$ continuous spin supermultiplet in 4-dimensional Minkowski space-time. We observe that, in the CSP supermultiplet, the bosonic field should be a complex scalar continuous spin field and the fermionic one must be a Dirac continuous spin field:
\be 
\hbox{$\mathcal{N}=1$ CSP supermultiplet} \qquad \Rightarrow \qquad \bigg(~\hbox{complex scalar CSP}~~,~~ \hbox{Dirac CSP}~\bigg)\label{CSP super}\,.
\ee

\vspace{.2cm}

We note that the first supersymmetry transformations, in the frame-like approach, for the $\mathcal{N}=1$ continuous (infinite) spin supermultiplet was presented by Zinoviev \cite{Zinoviev:2017rnj} in three-dimensional Minkowski space-time, which was recently generalized to four dimensions \cite{Buchbinder:2019kuh}, however our approach to supersymmetrize the theory in this paper is different. Furthermore, there are other papers discussing the supersymmetric continuous (infinite) spin gauge theory \cite{Buchbinder:2019iwi}-\cite{Buchbinder:2019sie}. 
Apart from supersymmetry, a number of papers have studied other aspects of the continuous spin theory in different approaches \cite{Khan:2004nj}-\cite{Burdik:2019tzg}. For instance, since an interacting theory is more favored, possible interactions of continuous spin particle with matter have been investigated in \cite{Metsaev:2017cuz,Bekaert:2017xin}, while interactions of continuous spin tachyon\,{\color{blue}\footnote{Also known as massive continuous spin particle.}} is examined in \cite{Rivelles:2018tpt,Metsaev:2018moa}.

\vspace{.2cm}

The presence of the dimensionful parameter $\m\neq0$ in the CSP theory makes it in some ways similar to a massive theory. More precisely, one may find an apparent connection (in formulations) between the massive higher spin gauge field theory and the continuous spin one. For instance, although continuous spin particle is massless, its representation is not conformally invariant since it is characterized by a parameter with the dimension of a mass, like massive particles \cite{Bekaert:2009pt}. Moreover, we shall show that the Dirac continuous spin field equation does not decouple into two Weyl equations, which is similar to the massive Dirac spin-$\frac{1}{2}$ field equation. In addition, the number of real CSP fields we use for the $\mathcal{N}=1$ CSP supermultiplet \eqref{CSP super} equals the number of real fields in the massive higher spin $\mathcal{N}=1$ supermultiplet, in which two bosonic fields (with opposite parity) and two fermionic fields are used \cite{Zinoviev:2007js}. On the other side, there is a tight connection between the massless higher spin field theory and the continuous spin one at $\m=0$ (refer e.g. to footnote \ref{foot}). These two connections with the massive and massless higher spin gauge theories can give us a better understanding of how to deal with and develop the continuous spin gauge field theory.


\vspace{.2cm}

The layout of this paper is as follows. In section \ref{HS review}, we will briefly review the supersymmetric higher spin theory \`a la Fronsdal for both half-integer and integer spin supermultiplets. The review contains and pursues a method we have used to find the supersymmetry transformations for the CSP theory, however, the reader can jump to next section and follows the main part of the paper. In section \ref{sec. unconstraint}, we will present supersymmetry transformations for unconstrained formulation, while in section \ref{sec. const}, we will provide those for the constrained formulation of the CSP theory. In section \ref{sec. rel}, we will make a connection between two obtained supersymmetry transformations, presented in \ref{sec. unconstraint} and \ref{sec. const}. The conclusions are displayed in section \ref{conclu}. In appendices; we present our conventions in the appendix \ref{conven}. The appendix \ref{proof} includes a proof related to section \ref{sec. unconstraint}. Transformation rules of the chiral supermultiplet will be presented in appendix \ref{Wess-Zu}. A discussion on inverse operators will be displayed in appendix \ref{I op}. Useful relations concerning supersymmetry and so on will be presented in the appendix \ref{usef}.

\section{SUSY higher spin gauge theory: a brief review}\label{HS review}

This section reviews the massless half-integer and integer higher spin $\mathcal{N}=1$ supermultiplets in four-dimensional Minkowski space-time in the metric-like approach which is known for a long time \cite{Curtright:1979uz}\,{\color{blue}\footnote{We note that a frame-like approach was given by Vasiliev in \cite{Vasiliev:1980as}. We note also that off-shell superfield realizations of $\mathcal{N}=1$, $d = 4$ higher superspin massless multiplets were given in \cite{Kuzenko:1993jp,Kuzenko:1993jq}.}}\,(see also \cite{Zinoviev:2007js} for review). However, our approach is based on the generating functions and we deal with operators, so as this fashion somewhat facilitates calculations. Moreover, the applied method to supersymmetrize the massless higher spin (HS) theory in this section has been employed for the CSP theory in sections \ref{sec. unconstraint} and \ref{sec. const}, so in this respect the present review may be informative.

\vspace{.2cm}

In 4-dimensional flat space-time, a real massless bosonic higher spin field (except the spin-$0$ field which has one degree of freedom) has two degrees of freedom. Thus one can consider a Majorana spinor as its superpartner, which has also two real degrees of freedom for any arbitrary half-integer spin. Therefore, in what follows, we will take into account the Fronsdal actions \cite{Fronsdal:1978rb} (except the Klein-Gordon action) for real massless higher spin fields, as well as the Fang-Fronsdal actions \cite{Fang:1978wz} in which the fermionic field is a Majorana spinor. Two possible supermultiplets, half-integer and integer ones, will be discussed separately in the following.

\subsection{Half-integer spin supermultiplet: {$(\,s\,,\, s\, {\scriptstyle +\, 1/2}\,)$}}

Let us first introduce the bosonic and fermionic massless higher spin fields, respectively, by the generating functions
\be 
\phi_s(x,\w) = \frac{1}{s!} ~\w^{\m_{1}}   \ldots \w^{\m_{s}}\, \phi_{\m_{1}\ldots\m_{s}}(x) \,,
\quad\quad\quad
\psi_s(x,\w) =  \frac{1}{s!} ~\w^{\m_{1}}   \ldots \w^{\m_{s}}\, \psi_{\m_{1}\ldots\m_{s}}(x)\,, \label{gen.}
\ee 
where $\phi_{\m_{1}\ldots\m_{s}}$ is a tensor field of integer spin $s$, and $\psi_{\m_{1}\ldots\m_{s}}$ is a spinor-tensor field of half-integer spin $s+\frac{1}{2}$. To ignore the chiral supermultiplet $(\,{\scriptstyle 0}\,,\, {\scriptstyle 1/2}\,)$ which is irrelevant for higher spins, we consider $s \geqslant 1$ in the half-integer spin supermultiplet $(\,s\,,\, s\, {\scriptstyle +\, 1/2}\,)$ and consequently in the gauge fields \eqref{gen.}. The generating functions \eqref{gen.} are considered to be double- and triple gamma-traceless, that is 
\be 
(\dww)^2\,\phi_s(x,\w)=0\,, \qquad\qquad (\dws)^3\, \psi_s(x,\w)=0\,,
\ee 
and obey the following homogeneity conditions 
\be 
(N - s\, )\,\phi_s(x,\w)=0\,, \qquad\qquad (N - s\, )\,\psi_s(x,\w)=0\,,
\ee
where $N=w\c\dw$\,. Then, the Fronsdal \cite{Fronsdal:1978rb} and Fang-Fronsdal \cite{Fang:1978wz} actions can be given respectively by\,{\color{blue}\footnote{We note that the spinor field $\psi_s$ in \eqref{action f} is considered to be a Majorana field, thus the overall factor of $\frac{1}{2}$ compared to the Fang-Fronsdal action \cite{Fang:1978wz} is usual for selfconjugate fields, introduced to ensure a consistent normalization of the field operators in quantum field theory.}}
\begin{align}
I_s^b&=\frac{1}{2}\int d^4 x ~\phi_s(x,\dw)~\mathbf{B}~\phi_s(x,\w)~\Big|_{\w=0} \label{action b}\,, \\[8pt]
I_s^f&=\frac{1}{2} \int d^4 x ~\overline{\psi}_s(x,\dw)~\mathbf{F}~\psi_s(x,\w)~\Big|_{\w=0} \label{action f}\,,
\end{align}
where the operators $\mathbf{B}$ and $\mathbf{F}$ are respectively the bosonic and fermionic operators, defined as
\begin{align}
\mathbf{B}&:=-\,\Box+\wdx\dwdx-\tfrac{1}{2}\,\wdx^2\dww -\tfrac{1}{2}\, \w^2\dwdx^2+\tfrac{1}{2}\,\w^2\,\Box\,\dww + \tfrac{1}{4}\,\w^2\wdx\dwdx\dww  \,,\label{K_0^b}\\[8pt]
\mathbf{F}&:=\,i\,\Big[\, \ds-\ws\dwdx-\wdx\dws + \ws\ds\dws + \tfrac{1}{2}\,\ws\wdx\dww +\tfrac{1}{2}\,\w^2\dwdx\dws -\tfrac{1}{4}\,\w^2\ds \dww\, \Big]\,.\label{K_0^f}
\end{align}
We note that the hermiticity of the actions \eqref{action b}, \eqref{action f} satisfy by 
\begin{align} 
(\mathbf{B})^\dagger&=\mathbf{B}\,, \qquad\qquad\qquad\qquad~\,~\,(\mathbf{F})^\dagger=\g^0\,\mathbf{F}\,\g^0 \,,\label{B , F}\\ [\phi_s(x,\w)]^\dagger&=\phi_s(x,\dw)\,,\qquad\qquad~
[\psi_s(x,\w)]^\dagger=\psi_s^\dagger(x,\dw)\,,
\end{align}
with respect to the following Hermitian conjugation rules
\be
(\p_x^{\,\a})^\dag:=\,-\,\p_x^{\,\a} \,,\qquad (\dw^{\,\a})^\dag :=-\, \w^{\,\a} \,,\qquad (\w^{\,\a})^\dag:=-\,\dw^{\,\a} \,.
\label{hermitian conjugates}
\ee

The bosonic \eqref{action b} and fermionic \eqref{action f} actions are invariant under the following gauge transformations
\begin{align}
\d\,\phi_s(x,\w)&=\wdx\,\xi_{s}(x,\w)\,,\\
\d\,\psi_s(x,\w)&=\wdx\,\zeta_{s}(x,\w)\,,
\end{align}
where $\xi_{s}$ and $\zeta_{s}$ are gauge transformation parameters introduced by the generating functions
\be 
\xi_s(x,\w) = \tfrac{1}{(s-1)!} ~\w^{\m_{1}}   \ldots \w^{\m_{s-1}}\, \xi_{\m_{1}\ldots\m_{s-1}}(x) \,,
\quad\quad\quad
\zeta_s(x,\w) =  \tfrac{1}{(s-1)!} ~\w^{\m_{1}}   \ldots \w^{\m_{s-1}}\, \zeta_{\m_{1}\ldots\m_{s-1}}(x)\,, \label{gen. 1}
\ee  
subject to the traceless and gamma-traceless conditions 
\be 
(\dww)\,\xi_s(x,\w)=0\,, \qquad\qquad \dws\, \zeta_s(x,\w)=0\,.
\ee 

In order to find supersymmetry transformations which leave invariant the sum of both free actions \eqref{action b}, \eqref{action f}
\be 
I_{(s,\,s+1/2)}=I_s^b + I_s^f\,,\label{susy action} \qquad\qquad ;\,s \geqslant 1
\ee 
one can consider the following ansatz
\begin{align}
\d \,\phi_s(x,\w) &=\a\, \overline{\epsilon}\,\psi_s(x,\w) \,, \qquad\qquad\qquad~~~\, \d \,\psi_s(x,\w)= {\mathbf{X}}\,\phi_s(x,\w)\,\epsilon\,,  \\
\d \,\phi_s(x,\dw) &=\a\,\overline{\psi}_s(x,\dw)\,\epsilon\,, \qquad\qquad~~~~~\,\,\, \d\, \overline{\psi}_s(x,\dw) = \overline{\epsilon}\,\phi_s(x,\dw)\,{\mathbf{X}}\,,
\end{align}
where $\epsilon$ is the global supersymmetry transformation parameter which is a Majorana spinor, $\a$ is considered to be a real number determining from the closure of the SUSY algebra, and ${\mathbf{X}}$ (assuming that ${\mathbf{X}}^\dagger=\g^0\,{\mathbf{X}}\,\g^0$) is an operator which we would like to find out. To this end, one can vary the SUSY action \eqref{susy action} with respect to the ansatz, which yields
\begin{align}
\d I_{(s,\, s+1/2)}=\d I_s^b + \d I_s^f
&=\frac{1}{2}\int d^4 x\le[\,\a\,\overline{\psi}_s\,\epsilon\,\mathbf{B}\,\phi_s+\phi_s\,\mathbf{B}\,\a\,\overline{\epsilon}\,\psi_s+ \overline{\epsilon}\,\phi_s\,{\mathbf{X}}\,\mathbf{F}\,\psi_s+\overline{\psi}_s\,\mathbf{F}\,{\mathbf{X}}\,\phi_s\,\epsilon \,\ri]~\Big|_{\w=0}\nonumber\\
&=\frac{1}{2}\int d^4 x\le[\,\overline{\psi}_s\le(\a\,\mathbf{B}+\mathbf{F}\,{\mathbf{X}}\ri)\phi_s\,\epsilon + \overline{\epsilon}\,\phi_s \le(\a\,\mathbf{B}+{\mathbf{X}}\,\mathbf{F}\ri)\psi_s~\ri]~\Big|_{\w=0}\,.\label{aa}
\end{align}
Demanding $\d I_{(s,\, s+1/2)} =0$, one can cancel the first term in \eqref{aa} by choosing
\be 
\a\,\mathbf{B}=-\,\mathbf{F}\,{\mathbf{X}} 
\,.\label{cond.}
\ee 
Then, taking hermitian conjugation of \eqref{cond.} leads to $\a\,\mathbf{B}=-\,{\mathbf{X}}\,\mathbf{F}$ which, in turn, vanishes the second term of \eqref{aa}. Now we are in a position to find the operator $\mathbf{X}$. For this purpose, we consider a general form (which is considered to be similar to the fermionic operator $\mathbf{F}$ \eqref{K_0^f}) with undetermined coefficients
\begin{align}
\mathbf{X}&=i\,\Big[\,\ds\, A_1+\ws\dwdx\, A_2+\wdx\dws\, A_3 + \ws\ds\dws\, A_4  + \ws\wdx\dww\, A_5 +\w^2\dwdx\dws\, A_6 +\w^2\ds \dww \,A_7\,\Big]\,, \label{X}
\end{align}
where $A_i$ ($i=1,\dots,7$) are considered to be real functions of $N (:=\w\c\dw)$ to satisfy our assumption: ${\mathbf{X}}^\dagger=\g^0\,{\mathbf{X}}\,\g^0$\,. Plugging the operators \eqref{K_0^b}, \eqref{K_0^f}, \eqref{X} into \eqref{cond.}, and using (anti-)commutation relations presented in appendix of \cite{Najafizadeh:2017acd} which lead to some useful relations \eqref{01}-\eqref{10}, one can read the coefficients $A_i$\,, which become
\be 
A_1=-\,\a\,, \quad\quad\quad\quad\quad A_2=-\,A_4=2A_5=\frac{\a}{2N}\,,  \quad\quad\quad\quad\quad A_3,\, A_6,\, A_7 =0\,.
\ee 
Therefore, we could determine the operator $\mathbf{X}$ and consequently the expression for $\d\,\psi_s$, which is
\begin{align}
\d \,\psi_s(x,\w)&=\mathbf{X} ~\phi_s(x,\w) \,\epsilon \\[5pt] &=-\,i\,\a\le[~\ds - \ws\,\tfrac{1}{2(N+1)}\,\dwdx+\ws\,\ds\,\tfrac{1}{2(N+1)}\,\dws - \ws\,\wdx\,\tfrac{1}{4(N+2)}\,\dww~\ri] \phi_s(x,\w) \,\epsilon\,.\nonumber
\end{align}
%
To find the parameter $\a$, we should check the closure of the SUSY algebra. We will then find that the algebra closes up to a field dependent gauge transformation parameter by choosing $\a=\sqrt{2}$, that is
\be 
[\,\d_1\,,\,\d_2\,]\,\phi_s(x,\w) =-\, 2\,i\,(\bar\epsilon_2\,\ds\,\epsilon_1)\phi_s(x,\w)+(\w\c\p_x)\,\xi_{s}(\phi)\,,
\ee 
where
\be 
\xi_{s}(\phi) = \tfrac{i}{N+1}\,\Big[(\bar\epsilon_2\,\ws\,\epsilon_1)\,\dww- 2\,(\bar\epsilon_2\,\ds_\w\,\epsilon_1) \Big]\phi_s(x,\w)\,.
\ee 
To illustrate how the SUSY algebra closes we have used the Majorana flip relations \eqref{maj flip} and the identity \eqref{123}. 

Hence, we find that the SUSY action \eqref{susy action} is invariant under the following supersymmetry transformations:
\begin{tcolorbox}[ams align, colback=white!100!black]
	\d\, \phi_s(x,\w)&={\scriptstyle\sqrt{2}}\,\,\overline{\epsilon}\,\psi_s(x,\w) \,,\label{hss1c1}\\
	\d \,\psi_s(x,\w) &=-\, \tfrac{i}{\sqrt{2}}\le[~2\,\ds - \ws\,\tfrac{1}{N+1}\,\dwdx+\ws\,\ds\,\tfrac{1}{N+1}\,\dws - \ws\,\wdx\,\tfrac{1}{2(N+2)}\,\dww~\ri] \phi_s(x,\w) \,\epsilon\,.\label{hss2c1}
\end{tcolorbox}
\hspace{-10pt}This is equivalent to the well-known result of the supersymmetry transformations for the half-integer spin supermultiplets $(\,s\,,\, s\, {\scriptstyle +\, 1/2}\,)$ with $s\geqslant 1$, which was first presented by Curtright in \cite{Curtright:1979uz}.


\subsection{Integer spin supermultiplet: {$(\, s\, {\scriptstyle +\, 1/2}\,,\,s\,{\scriptstyle +\,1}\,)$}}

Let us take into account $s \geqslant 0$ for the integer spin supermultiplet $(\, s\, {\scriptstyle +\, 1/2}\,,\,s\,{\scriptstyle +\,1}\,)$, and as a result for the generating functions in \eqref{gen.}. In this case, one can consider the bosonic \cite{Fronsdal:1978rb} and fermionic \cite{Fang:1978wz} higher spin actions respectively by
\begin{align}
I_{s+1}^b&=\frac{1}{2}\int d^4 x ~\phi_{s+1}(x,\dw)~\mathbf{B}~\phi_{s+1}(x,\w)~\Big|_{\w=0} \label{action b 1}\,,\\[8pt] 
I_s^f&=\frac{1}{2} \int d^4 x ~\overline{\psi}_s(x,\dw)~\mathbf{F}~\psi_s(x,\w)~\Big|_{\w=0} \label{action f 1}\,,
\end{align}
where the bosonic $\mathbf{B}$ and fermionic $\mathbf{F}$ higher spin operators were given in \eqref{K_0^b}, \eqref{K_0^f}\,.
In order to find the supersymmetry transformations for the SUSY action
\be 
I_{(s+1/2,\,s+1)}=I_{s+1}^b+ I_s^f\,,\label{susy action 2} \qquad\qquad;\,s\geqslant 0
\ee 
one can start with the following ansatz
\begin{align}
\d \,\phi_{s+1}(x,\w) &=i\, \overline{\epsilon}\,\ws\,f\,\psi_s(x,\w) \,, \qquad\qquad\qquad~~~~~ \d \,\psi_s(x,\w)= {{\mathbf{Y}(\dw)}}\,\phi_{s+1}(x,\w)\,\epsilon\,,  \\\d \,\phi_{s+1}(x,\dw) &=i\,\overline{\psi}_s(x,\dw)\,f\,\dws\,\epsilon
\,, ~~~~~~~~~\,\,\quad~~~~\,\,\, \d\, \overline{\psi}_s(x,\dw) \,=-\, \overline{\epsilon}\,\phi_{s+1}(x,\dw)\,{\mathbf{Y}(\w)}\,,
\end{align}
where $\epsilon$ is the global supersymmetry transformation parameter, $f$ can be in general a real function of $N$ determining from the closure of the SUSY algebra, and $\mathbf{Y}$ is an operator (assuming that \,${\mathbf{Y}(\dw)}^\dagger=-\,\g^0\,\mathbf{Y}(\w)\,\g^0$\,) that we would like to find. We note that presence of the unit imaginary number $i$ in the ansatz guaranties that the Majorana spinor field is real. Varying the SUSY action \eqref{susy action 2} with respect to the above ansatz, we will reach to
\begin{align}
\d I_{(s+1/2 ,\, s+1)}&=\d I_{s+1}^b+\d I_s^f \nonumber\\[6pt]
&=\frac{1}{2}\int d^4 x\le[\,i\,\overline{\psi}_s\,f\,\dws\,\epsilon\,\mathbf{B}\,\phi_{s+1}\,+\phi_{s+1}\,\mathbf{B}\,i\,\overline{\epsilon}\,\ws\,f\,\psi_s\,-\,\overline{\epsilon}\,\phi_{s+1}\,\mathbf{Y}(\w)\,\mathbf{F}\,{\psi}_s+\overline{\psi}_s\,\mathbf{F}\,\,{\mathbf{Y}(\dw)}\,\phi_{s+1}\,\epsilon\,\ri]\nonumber\\[6pt]
&=\frac{1}{2}\int d^4 x\le[\,i\,\overline{\psi}_s\Big[\,f\,\dws\,\mathbf{B}-i\,\mathbf{F}\,{\mathbf{Y}(\dw)}\,\Big]\phi_{s+1}\,\epsilon +i\,\overline{\epsilon}\,\phi_{s+1}\Big[\,\mathbf{B}\,\ws\,f+i\,\mathbf{Y}(\w)\,\mathbf{F}\,\Big]\psi_s\ri]\,.
\end{align}
Demanding $\d I_{(s+1/2,\,s+1)} =0$, we have to choose
\be 
f\,\dws\,\mathbf{B}=i\,\mathbf{F}\,{\mathbf{Y}(\dw)}\,,  \label{cond. 2}
\ee 
leading in turn to $\mathbf{B}\,\ws\,f=-\,i\,\mathbf{Y}(\w)\,\mathbf{F}$, by taking hermitian conjugation of \eqref{cond. 2}. Hence, the remaining task is determining the operator $\mathbf{Y}(\dw)$. Considering the property we adopted to the operator $\mathbf{Y}(\dw)$, one can drop an $\ws$ from the left-hand-side of the operator $\mathbf{F}$ \eqref{K_0^f}, and surmise a general form for $\mathbf{Y}(\dw)$ as
\begin{align}
{\mathbf{Y}(\dw)}&= B_1\,\dwdx +B_2\,\, \ds\,\dws +B_3\, \wdx\dww  +B_4\,\,\ws\dwdx\dws  +B_5\,\,\ws\ds \dww\,, \label{Y-0}
\end{align}
where $B_i$ ($i=1,\dots,5$) are considered to be real functions of $N (:=\w\c\dw)$. Then, plugging \eqref{K_0^b}, \eqref{K_0^f}, \eqref{Y-0} into \eqref{cond. 2}, and applying the (anti-)commutation relations presented in appendix of \cite{Najafizadeh:2017acd}, we will find the coefficients as
\be 
B_1=-\,f\,, \quad\quad\quad B_2=f\quad\quad\quad  B_3=-\,\tfrac{1}{2}\,f \quad\quad\quad\quad\quad B_4,B_5=0\,.
\ee 
Therefore, the operator $\mathbf{Y}(\dw)$ and as a result the expression for $\d\psi_s$ can be given by
\begin{align} 
\d \psi_s(x,\w)&=\mathbf{Y} (\dw)\,\phi_{s+1}(x,\w)\,\epsilon \\[5pt] &=f\,\le[~\ds\,\dws-\dwdx- \tfrac{1}{2}\,\wdx\,\dww~\ri] \phi_{s+1}(x,\w) \,\epsilon\,.\nonumber
\end{align}
The closure of the SUSY algebra will fix the $f$ operator. In fact, we will find by choosing 
\be 
f=\frac{1}{\sqrt{{N+1}}}\,,
\ee 
the algebra will be closed up to a field dependent gauge transformation parameter
\be 
[\,\d_1\,,\,\d_2\,]\,\phi_{s+1} =-\, 2\,i\,(\bar\epsilon_2\,\ds\,\epsilon_1)\,\phi_{s+1}+(\w\c\p_x)\,\xi_{s+1}(\phi)
\ee 
where
\be 
\xi_{s+1}(\phi) =-\, \frac{i}{\sqrt{{N+1}}}\,\Big[(\bar\epsilon_2\,\ws\,\epsilon_1)\,\dww- 2\,(\bar\epsilon_2\,\ds_\w\,\epsilon_1) \Big]\phi_{s+1}\,.
\ee 
Therefore, we find that the SUSY action \eqref{susy action 2} is invariant under the following supersymmetry transformations:
\begin{tcolorbox}[ams align, colback=white!100!black]
	\d\, \phi_{s+1}(x,\w) &=i\, \overline{\epsilon}\,\ws\,\tfrac{1}{\sqrt{{N+1}}}\,\psi_s(x,\w) \,,\label{hss1c}\\
	\d\, \psi_s(x,\w)&=\tfrac{1}{\sqrt{{N+1}}}\le[~\ds\,\dws-\dwdx-\tfrac{1}{2}\,\wdx\,\dww~\ri]\,\phi_{s+1}(x,\w)\,\epsilon\,.\label{hss2c}
\end{tcolorbox}
\hspace{-10pt}This is also equivalent to the well-known result of the supersymmetry transformations for the integer spin supermultiplets $(\, s\, {\scriptstyle +\, 1/2}\,,\,s\,{\scriptstyle +\,1}\,)$ with $s\geqslant 0$, which was first discovered in \cite{Curtright:1979uz}.

\section{Unconstrained formulation of the CSP theory} \label{sec. unconstraint}

This section, and the next one, include main results of this paper. As we know, a general property of all supersymmetric theories is that the number of physical bosonic degrees of freedom is always identical to the number of fermions. On the other hand, we know that a continuous spin particle (bosonic or fermionic) has infinite number of physical degrees of freedom per space-time point. Hence, the equality of the number of bosonic and fermionic degrees of freedom in a CSP supermultiplet looks like meaningless. Therefore, in four-dimensional flat space-time, there would be in principle four possibilities for the $\mathcal{N}=1$ supermultiplet containing of a CSP and CSPino (superpartner of CSP)
\be 
\hbox{$\mathcal{N}=1$ CSP supermultiplet} \qquad \Rightarrow \qquad \Big(~\hbox{CSP}~~,~~ \hbox{CSPino}~\Big) \label{cspinp}
\ee
so as one can consider CSP to be a real or complex scalar continuous spin field, while one may consider CSPino to be a Majorana or Dirac continuous spin field. Among these possibilities, we find that the mentioned case in \eqref{CSP super} with complex scalar CSP field and Dirac CSP field is the only choice which is consistent with supersymmetry expectations.

Here, in this section, we first present bosonic \cite{Schuster:2014hca} and fermionic \cite{Najafizadeh:2015uxa} unconstrained formulations of the continuous spin gauge field theory. Then we provide supersymmetry transformations for the $\mathcal{N}=1$ continuous spin supermultiplet which leave the sum of the bosonic and fermionic actions invariant and simultaneously satisfy the SUSY algebra, as we expect. We also investigate the helicity limit of the SUSY CSP theory and supersymmetrize unconstrained formulation of the higher spin gauge field theory \`a la Segal, given by the actions \cite{Segal:2001qq}, \cite{Najafizadeh:2018cpu}. We note that this formulation of the higher spin theory has not been already supersymmetrized. Notice again that in these formulations of the CSP and HS theories there is no constraint on gauge fields (bosonic or fermionic) and their related gauge transformation parameters.

\subsection{Bosonic action}
Let us consider the Schuster-Toro action \cite{Schuster:2014hca} in four-dimensional Minkowski space-time, in which the scalar continuous spin gauge field is complex. Applying partial integration to the Schuster-Toro action, the complex scalar continuous spin gauge field action is given by\,{\color{blue}\footnote{We note an overall factor of $\tfrac{1}{2}$ has been dropped from the Schuster-Toro action \cite{Schuster:2014hca} compared to \eqref{S-T action}, because we deal with a complex scalar CSP field.}}
\begin{align}
S_{_{CSP}}^{\,b}&= \int d^4x\, d^4\e\,\,\delta'(\e^2+1)\,\Phi^\dagger(x,\e)\,\Big[-\,\Box + (\e\c\p_x)(\de\c\p_x+\m\,)-\tfrac{1}{2}\,(\e^2+1)(\de\c\p_x+\m\,)^2\Big]\,\Phi(x,\e) \,,\label{S-T action}
\end{align}
where $\m$ is continuous spin parameter, $\e^{\,\m}$ is a 4-dimensional auxiliary Lorentz vector localized to the unit hyperboloid $\e^2=-1$, and $\d'$
is the derivative of the Dirac delta function with respect to its argument, i.e. $\delta'(a)=\frac{d}{da}\,\delta(a)$\,. The complex scalar CSP field $\Phi$ is unconstrained and introduces by a collection of totally symmetric complex tensor
fields $\Phi_{\m_1 \dots \m_s}(x)$ of all integer rank $s$, packed into a single generating function 
\be 
\Phi(x,\e)=\sum_{s=0}^{\infty}\,\frac{1}{s!}~\e^{\m_1} \dots \e^{\m_s}~\Phi_{\m_1 \dots \m_s}(x)\,. \label{phi}
\ee 
We note that in the infinite tower of spins \eqref{phi},
every spin state interns only once, and the spin states are mixed under the Lorentz boost, so as the degree of mixing is controlled by the continuous spin parameter $\m$.  

The action \eqref{S-T action} is invariant under gauge transformations 
\begin{align}
\delta_{\xi_1} \Phi (x,\e)&= \le[    \,\e \cdot \p_x  -  \tfrac{1}{2}\, (\e^2+1  )   (\p_\e \cdot \p_x  +\m\,  ) \ri] \xi_1 (x,\e)\,,\label{gt b 1} \\
\delta_{\xi_2} \Phi (x,\e)&=(\e^2 + 1 )^2 \, {\xi_2}(x,\e)\,,\label{gt b 2}
\end{align}
where $\xi_1, \xi_2$ are two arbitrary complex gauge transformation parameters, which are unconstrained. By varying the action \eqref{S-T action} with respect to the gauge fields $\Phi^\dagger$ and $\Phi$, one can obtain two independent equations of motion which the one for the CSP gauge field $\Phi$ reads
\be
\delta'(\e^2+1)\le[\,-\,\Box + (\e\c\p_x)(\de\c\p_x+\m\,)-\tfrac{1}{2}\,(\e^2+1)(\de\c\p_x+\m\,)^2\ri]\Phi(x,\e)=0\,,\label{bcsp eom}
\ee  
besides a same independent equation of motion for the CSP gauge field $\Phi^\dagger$.

\subsection{Fermionic action}
Let us now consider the fermionic version of the Schuster-Toro's action in four-dimensional flat space-time \cite{Najafizadeh:2015uxa}, in which the fermionic continuous spin field is a Dirac spinor. By applying partial integration to the action \cite{Najafizadeh:2015uxa}, the Dirac continuous spin gauge field action is given by   
\begin{align}
S_{_{CSP}}^{\,f}&=
\int d^4x\, d^4\e~\delta'(\e^2+1)\,\overline\Psi(x,\e)\,(\es+i\,)\,\Big[\,\ds - (\es-i\,)(\de\c\p_x+\m\,)\,\Big]\,\Psi(x,\e)\,,\label{Mojtaba}
\end{align}
where $\ds$ (or $\es$) is defined according to the Feynman slash notation: $\ds\equiv\g^\m\,\p_\m$ with $\g^\m$s as the gamma matrices in 4 dimensions\,. The fermionic SCP field $\Psi$ is considered to be a Dirac spinor field, which is unconstrained and introduced by the generating function 
\be 
\Psi(x,\e)=\sum_{s=0}^{\infty}\,\frac{1}{s!}~\e^{\m_1} \dots \e^{\m_s}~\Psi_{\m_1 \dots \m_s}(x)\,, \label{psi}
\ee 
where $\Psi_{\m_1 \dots \m_s}(x)$ are totally symmetric Dirac spinor-tensor fields of all half-integer spin $s+\frac{1}{2}$\,, in such a way that the spinor index is left implicit\,. Again, as the bosonic case, in the infinite tower of spins \eqref{psi}, every spin state interns only once, and the spin states mix under the Lorentz boost which the degree of mixing is controlled by the continuous parameter $\m$.

The action \eqref{Mojtaba} is invariant under spinor gauge transformations
\begin{align}
\delta_{\zeta_1} \,\Psi(x,\e)&=\le[\,\ds \,(\es - i \,) - (\eta^2 +1 ) ({\partial_{\eta}} \cdot \p_x +\m\,) \,\ri]  {\zeta_1}(x,\e)\,, \label{zet 1}\\
\delta_{\zeta_2} \,\Psi(x,\e)&= (\e^2+1) ( \es +  i\,) \, {\zeta_2}(x,\e)\,,\label{zet 2}
\end{align}
where $\zeta_1$, $\zeta_2$ are the unconstrained arbitrary spinor gauge transformation parameters\,.

Varying the action \eqref{Mojtaba} with respect to the spinor gauge field $\overline\Psi$ yields the equation of motion for the unconstrained Dirac continuous spin field $\Psi$
\be 
\delta'(\e^2+1)\,(\es+i\,)\le[\,\ds - (\es-i\,)(\de\c\p_x+\m\,)\,\ri]\Psi(x,\e)=0\,. \label{EOM f}
\ee

We note that there are two possibilities for presenting the unconstrained formulation of the fermionic CSP/HS theory (see appendix of B in \cite{Bekaert:2017xin}). One possibility is the one we have used in \cite{Najafizadeh:2018cpu} and here. Another possibility can be expressed by converting $i \rightarrow -\,i$ in relations of \eqref{Mojtaba}-\eqref{EOM f} which have been used in \cite{Najafizadeh:2015uxa}, \cite{Bekaert:2017xin}. 


\vspace{.5cm}

\noindent\textbf{Remark:} 

\vspace{.2cm}

\noindent Here, we recall that a four-component Dirac spinor field can be written as
$$\psi=\le( \begin{array}{c}
\psi_L  \\
 \psi_R \\
\end{array} \ri)\,,$$
where the two-component objects $\psi_L$ and $\psi_R$ are left-handed and right-handed Weyl spinors respectively. If one uses notation in \cite{Peskin:1995ev} which defines
\be
\s^\m:=(\,1\,,\,\vec{\s}\,)\,,\qquad\qquad {\bar\s}^\m:=(\,1\,,\,-\,\vec{\s}\,)\,,\qquad\qquad \hbox{so that}~~~\qquad\g^{\m}=\le( \begin{array}{cc}
	0 & \s^\m \\
	\bar{\s}^\m & 0 \\
\end{array} \ri)\,,
\ee 
then the Dirac equation for massive spin $\tfrac{1}{2}$ particle can be written as
\be 
\le( \begin{array}{cc}
	-\,m & i\,\s\c\p \\
	i\,{\bar\s}\c\p & -\,m \\
\end{array} \ri)\le( \begin{array}{c}
\psi_L  \\
\psi_R \\
\end{array} \ri)=0\,,
\ee 
demonstrating the two Lorentz group representations $\psi_L$ and $\psi_R$ are mixed by the mass term in the Dirac equation. However, in massless case, the equations for $\psi_L$ and $\psi_R$ decouple and yield Weyl equations:
\be 
i\,{\bar\s}\c\p\,\psi_L=0\,,\qquad\qquad i\,{\s}\c\p\,\psi_R=0\,.
\ee 

Based upon the above discussion, as the Dirac continuous spin gauge field (given by the equation of motion \eqref{EOM f}) describes a massless particle, one can expect to derive the so-called Weyl continuous spin equations. To this end, using the above notation, let us write \eqref{EOM f} in terms of $\Psi_L$ and $\Psi_R$ 
\be 
\delta'(\e^2+1) \le( \begin{array}{cc}
	\mathcal{M} & i\,\s\c\p \\
	i\,{\bar\s}\c\p &  {\mathcal{\overline M}}\\
\end{array} \ri) \le( \begin{array}{c}
	\Psi_L  \\
	\Psi_R \\
\end{array} \ri)=0\,,\label{Weyl CSP}
\ee 
where
\be
\mathcal{M}:=(\s\c\e)(\bar\s\c\p)-(\e^2+1)(\de\c\p_x+\m)\,,\quad\quad\quad
{\mathcal{\overline M}}:=(\bar\s\c\e)(\s\c\p)-(\e^2+1)(\de\c\p_x+\m)\,.
\ee  
This equation, manifestly, demonstrates that the Dirac continuous spin equation \eqref{Weyl CSP} can not be decoupled into two independent Weyl CSP equations. Even in the helicity limit ($\m \rightarrow 0$) which massless higher spin equations are expected to be reproduced, the equation \eqref{Weyl CSP} does not decompose into Weyl equations. However, the latter case happens due to the unconstrained formulation we use, so as in the constrained formulation (next section) we will see it can be decomposed.

\subsection{Supersymmetry transformations}\label{sec. uncon}

Now we are in a position to supersymmetrize unconstrained formulation of the continuous spin theory in 4-dimensional flat space-time for the $\mathcal{N}=1$ supermultiplet, in which we have considered the bosonic CSP as a complex scalar continuous spin filed and the fermionic CSP as a Dirac continuous spin field. By this feature, we find conveniently that the SUSY CSP action, a sum of the bosonic \eqref{S-T action} and fermionic \eqref{Mojtaba} CSP actions
\be 
S_{_{CSP}}^{^{\,SUSY}}=S_{_{CSP}}^{\,b}+S_{_{CSP}}^{\,f}\label{48}
\ee 
is invariant under the following supersymmetry transformations
\vs{.1cm}
\begin{tcolorbox}[ams align, colback=white!98!black]
	&\delta\,\Phi(x,\e)\, =\,\tfrac{1}{\sqrt{2}}\,\, \bar\epsilon\,\,\big(\,1+\g^5\,\big)\,\big(\es-i\,\big)\,\Psi(x,\e)\,, \label{hss1}\\[8pt]
	&\delta\,\Psi(x,\e)\, =\,\tfrac{1}{\sqrt{2}}\, \le[\,\ds-\tfrac{1}{2}\,\big(\es+i\,\big)\big(\de\c\p_x+\m\,\big)\ri]\big(\,1-\g^5\,\big)\,\epsilon~\Phi(x,\e)\,,\label{hss2}
\end{tcolorbox}
\hspace{-.5cm}
where $\epsilon$ is an arbitrary constant\,{\color{blue}\footnote{We discuss global supersymmetry which means that $\epsilon$ is a constant, satisfying $\p_\m\,\epsilon=0$.}} infinitesimal, anticommuting, Dirac spinor object that parameterizes the supersymmetry transformations (see \eqref{maj flip} for its properties), $\g^5$ is the fifth gamma matrix, and $\Phi$, $\Psi$ are respectively the complex scalar and Dirac CSP fields.

Let us now calculate commutator of the supersymmetry transformations \eqref{hss1}, \eqref{hss2} acting on the bosonic and fermionic CSP fields. We straightforwardly find the SUSY commutator on the bosonic CSP field yields 
\be 
[\,\d_1\,,\,\d_2\,]\,\Phi(x,\e)=-\,2\,i\,(\bar\epsilon_2\,\ds\,\epsilon_1)\,\Phi(x,\e)\,,\label{b clou}
\ee 
which corresponds to the translation, while the one on the fermionic CSP field becomes
\begin{align} 
[\,\d_1\,,\,\d_2\,]\,\Psi(x,\e)\approx&\,-\,2\,i\,(\bar\epsilon_2\,\ds\,\epsilon_1)\,\Psi(x,\e) \label{susy f}\\[5pt]
&\,+\,\Big[\,\ds \,(\es - i \,) - (\eta^2 +1 ) ({\partial_{\eta}} \cdot \p_x +\m\,) \,\Big] \Big[\,\tfrac{1}{2}\,\bar\epsilon_1\,\g_\m\,\epsilon_2\,\g^\m\,(1-\g^5\,)\,\Psi(x,\e)\,\Big]\,.\nonumber
\end{align}
where `` $\approx$ '' denotes that we have applied the Dirac continuous spin field equation of motion \eqref{EOM f}. Taking into account a field dependent spinor gauge transformation parameter, given by
\be 
\zeta_1(\Psi)=\Big[\,\tfrac{1}{2}\,\bar\epsilon_1\,\g_\m\,\epsilon_2\,\g^\m\,(1-\g^5\,)\,\Psi(x,\e)\,\Big]\,,\label{f clou}
\ee 
the second line in \eqref{susy f} would be the $\zeta_1$ gauge transformation \eqref{zet 1}, demonstrating that the SUSY commutator acting on the fermion CSP field is closed on-shell, up to a gauge transformation. 
\vspace{.5cm}

\noindent\textbf{Remarks:} Concerning the supersymmetry transformations we obtained in \eqref{hss1} and \eqref{hss2}, there are some remarks which are useful to discuss: 

\vspace{.2cm}


\begin{itemize}
	
	\item By starting from the ansatz $\d\Phi=\a\, \bar\epsilon \, \Psi$ which $\a$ is an arbitrary parameter, one can prove\,{\color{blue}\footnote{We thank Mohammad Khorrami for discussion and sending us the proof.}} that there would be no $\d \Psi$ to leave invariant the sum of the bosonic and fermionic actions. In other words, in \eqref{hss1}, the term $(\es-i)$ is necessary for invariance of the SUSY action (see the appendix \ref{proof} for the proof).
	
	\item Employing the gamma fifth matrix $\g^5$ in the above set was essential for the closure of the SUSY algebra. In fact, by omitting $\g^5$, one can consider the bosonic field as a real scalar CSP field and the fermionic one as a Majorana or Dirac CSP field. However, in these two cases, although the SUSY action will be invariant under such transformations, the SUSY algebra will not be closed. 
	
	\item When the gamma fifth was employed, the CSP field has to be complex while the CSPino can be either a Majorana or a Dirac CSP field. We note that again a CSP has infinite physical degrees of freedom per space-time point, so as the Majorana or Dirac CSP field can be candidate of the CSPino. 
	
	\item If one chooses the Majorana CSP field as superpartner of the complex scalar CSP field, then the right-hand-side of \eqref{hss2} does not satisfy the Majorana spinor condition and should be improved by adding a complex conjugate of the right-hand-side of \eqref{hss2}. However, by adding the complex conjugate term, one finds that the SUSY algebra can not be closed\,{\color{blue}\footnote{We thank Dmitri Sorokin for many fruitful discussions on this issue.}}.

	\item For the $\mathcal{N}=1$ supermultiplet, we had to pick the Dirac CSP field as superpartner of the complex scalar CSP field. Therefore, one concludes that in the context of the CSP theory, instead of the equality of the number of bosonic and fermionic physical degrees of freedom, the number of bosonic and fermionic real CSP fields should be equal. This fact can be seen here for the $\mathcal{N}=1$ supermultiplet, and may hold for $\mathcal{N}>1$ but it remains to be checked.
	
	\item As we employed the Dirac CSP field, we deal with supersymmetry transformation parameter $\epsilon$ which is also a Dirac spinor object. Therefore, to illustrate how the SUSY algebra closes, we have used the so-called ``Dirac flip relations'' \eqref{maj flip}. Indeed, it is notable to mention that the Majorana flip relations hold also for Dirac spinors\,{\color{blue}\footnote{We thank Antoine Van Proeyen for clarifying the subject.}}(see \cite{Freedman:2012zz}, page 49, for more details)\,. 
	
	\item Although the supersymmetry transformation parameter $\epsilon$ is a Dirac spinor object, it is effectively not a Dirac spinor, but the right-handed Weyl spinor. This is due to the fact that, by defining $\frac{1}{2}\,(1-\g^5)\,\epsilon:=R\,\epsilon:=\epsilon_R$, one can see that $\epsilon$ always appears as $\overline{\epsilon_R}$ and $\epsilon_R$ in the SUSY transformations \eqref{hss1} and \eqref{hss2} respectively, which reflects the fact that we deal with $\mathcal{N}=1$ SUSY\,{\color{blue}\footnote{We thank again Dmitri Sorokin for pointing out this important comment.}}.


	\item In the frame-like approach, authors of \cite{Buchbinder:2019kuh} have supersymmetrized the infinite spin theory using four fields; two real infinite spin fields with opposite parity, as well as two Majorana infinite spin fields. In this regard, the number of real CSP fields we have used is quite consistent with \cite{Buchbinder:2019kuh}.

\end{itemize}

\subsection{Helicity limit}
By the term ``helicity limit''\,{\color{blue}\footnote{Authors of \cite{Schuster:2014hca} used the term ``helicity correspondence''.}}, we refer to a case that the continuous spin parameter $\m$ vanishes, and consequently the known results of the higher spin theory are expected to be reproduced. Since, in this section, we deal with unconstrained formulation of the CSP theory, it is natural to arrive at unconstrained formulation of the higher spin theory\,{\color{blue}\footnote{Here, by the term ``unconstrained'' we mean there are no constraints on the gauge fields and parameters, as well as there are no auxiliary fields in the theory. However, there are some differences in the meaning of the term, e.g. see unconstrained formulations in \cite{Francia:2002aa,Francia:2002pt,Francia:2007qt,Buchbinder:2007ak,Buchbinder:2008ss,Francia:2012rg,Francia:2013sca}.}} in the helicity limit. In the approach we follow here, unconstrained formulation of the bosonic higher spin gauge field theory was established by Segal in $d$-dimensional (A)dS space-time \cite{Segal:2001qq}. In four-dimensional flat space-time, this theory becomes the helicity limit of the Schuster-Toro formulation (see \cite{Schuster:2014hca}, \cite{Bekaert:2017xin} and \cite{Najafizadeh:2018cpu} for more details). In addition, unconstrained formulation of the fermionic higher spin gauge field theory was constructed in $d$-dimensional (A)dS space-time \cite{Najafizadeh:2018cpu}, which in four-dimensional flat space-time becomes the helicity limit of the fermionic CSP action \cite{Najafizadeh:2015uxa}. However, as we know, unconstrained formulation of the higher spin gauge theory \`a la Segal has not been supersymmetrized by now. Therefore, in the helicity limit, we will reach to a result that has not been already in the literature and thus its accuracy should be examined, what we will do here.   

At $\m=0$, the bosonic and frmionic CSP actions \eqref{S-T action}, \eqref{Mojtaba} reduce respectively to the following bosonic and fermionic higher spin actions
\begin{align}
S_{_{HS}}^{\,b}&= \int d^4x\, d^4\e\,\,\delta'(\e^2+1)\,\Phi^\dagger(x,\e)\,\Big[-\,\Box + (\e\c\p_x)(\de\c\p_x)-\tfrac{1}{2}\,(\e^2+1)(\de\c\p_x)^2\Big]\,\Phi(x,\e) \,,\label{Segal action}\\
S_{_{HS}}^{\,f}&=\int d^4x\, d^4\e~\delta'(\e^2+1)\,\overline\Psi\,(x,\e)\,\,\Big[\,(\es+i\,)\,\ds - (\e^2+1\,)(\de\c\p_x)\,\Big]\,\Psi(x,\e)\,,\label{Mojtaba hs}
\end{align}
where, here, the bosonic field $\Phi$ is a complex higher spin field and the fermionic one $\Psi$ is a Dirac higher spin field. These fields can be introduced respectively by the generating functions in \eqref{phi} and \eqref{psi}, but by this difference that here the infinite towers of spins are a direct sum over all integer helicity states ($s=0,1,\cdots,\infty$) and all half-integer helicity states ($s=1/2,3/2,\cdots,\infty$), in which helicity states do not mix under the Lorentz boost. These higher spin actions are invariant under gauge transformations \eqref{gt b 1}, \eqref{gt b 2}, \eqref{zet 1}, \eqref{zet 2}, and their equations of motion are given by \eqref{bcsp eom}, \eqref{EOM f}, when we set $\m=0$.  

By taking the helicity limit of the CSP theory (setting $\m=0$), one can propose that the SUSY higher spin action \`a la Segal, a sum of the complex higher spin action \eqref{Segal action} and the Dirac higher spin action \eqref{Mojtaba hs}
\be 
S_{_{HS}}^{^{\,SUSY}}=S_{_{HS}}^{\,b}+S_{_{HS}}^{\,f}\label{SUSY HS}\,,
\ee 
is invariant under the following supersymmetry transformations
\vs{.1cm}
\begin{tcolorbox}[ams align, colback=white!98!black]
	&\delta\,\Phi(x,\e)\, =\,\tfrac{1}{\sqrt{2}}\,\, \bar\epsilon\,\,\big(\,1+\g^5\,\big)\,\big(\es-i\,\big)\,\Psi(x,\e)\,, \label{S t 1}\\[8pt]
	&\delta\,\Psi(x,\e)\, =\,\tfrac{1}{\sqrt{2}}\, \le[\,\ds-\tfrac{1}{2}\,\big(\es+i\,\big)\big(\de\c\p_x\big)\ri]\big(\,1-\g^5\,\big)\,\epsilon~\Phi(x,\e)\,.\label{S t 2}
\end{tcolorbox}
\hspace{-.5cm}
We have examined and found that indeed the SUSY higher spin action \eqref{SUSY HS} is invariant under the above supersymmetry transformations, and the SUSY algebra closes on-shell up to a gauge transformation. More precisely, relations of \eqref{b clou}-\eqref{f clou} with $\m=0$ will be obtained for the closure of the SUSY higher spin algebra. Here, we just provided the supersymmetry transformations \eqref{S t 1}, \eqref{S t 2} for unconstrained formulation of the higher spin gauge theory (\`a la Segal), and let us postpone further discussion to subsection \ref{hl}, where we will investigate the helicity limit of constrained formalism.

\section{Constrained formulation of the CSP theory} \label{sec. const}

In this section, we first display bosonic \cite{Metsaev:2016lhs} and fermionic \cite{Metsaev:2017ytk} constrained formulations of the continuous spin gauge field theory in 4-dimensional flat space-time, discovered by Metsaev in $d$-dimensional (A)dS space-time. Then we provide supersymmetry transformations for the $\mathcal{N}=1$ continuous spin supermultiplet which leave the sum of the bosonic and fermionic actions invariant. Again, as the previous section, we consider a supermultiplet consist of one complex scalar CSP field and one Dirac CSP field.

\subsection{Bosonic action}

Let us define the complex scalar continuous spin gauge field as the generating function
\be 
\Phi(x,\w)=\sum_{s=0}^{\infty} ~ \frac{1}{s!} \,\w^{\m_{1}}   \ldots \w^{\m_{s}}\, \Phi_{\m_{1}\ldots\m_{s}}(x)\,, 
\ee
where $\Phi_{\m_{1}\ldots\m_{s}}$ represent for all totally symmetric complex tensor fields of all integer rank $s$, and $\w^\m$ is a 4-dimensional auxiliary vector. Then, the bosonic CSP action \cite{Metsaev:2016lhs}, in which the boson field is complex, is given by the complex scalar continuous spin action
\be 
I_{_{CSP}}^{\,b}=\int d^4 x~\Phi^\dagger(x,\dw)~\big(\,\mathbf{B}+\mathbf{B}_1+\mathbf{B}_2\,\big)~\Phi(x,\w)~\Big|_{\w=0}\,,\label{Metsaev b}
\ee
with
\bea
\hspace{-1cm}&&\mathbf{B}_{~}:=-\,\Box+\wdx\dwdx-\tfrac{1}{2}\,\wdx^2\dww -\tfrac{1}{2}\, \w^2\dwdx^2+\tfrac{1}{2}\,\w^2\,\Box\,\dww + \tfrac{1}{4}\,\w^2\wdx\dwdx\dww \,,\label{B}\\[8pt]
\hspace{-1cm}&&\mathbf{B}_1:=\,\m\,\Big[ \le(\w\c\p_x-\w^2\dwdx+\tfrac{1}{4}\,\w^2\wdx\dww\ri)\tfrac{-\,1}{\sqrt{2(N+1)}}+\tfrac{-\,1}{\sqrt{2(N+1)}} \le(\dw\c\p_x-\wdx\dww+\tfrac{1}{4}\,\w^2\dwdx\dww\ri)\Big] \label{B1}\\[5pt]
\hspace{-1cm}&&\mathbf{B}_2:=\m^2\Big[\,\tfrac{1}{2(N+1)}+\w^2\,\tfrac{1}{8(N+3)}\,\dww-\,\tfrac{1}{4}\,\w^2\,\tfrac{1}{\sqrt{(N+1)(N+2)}}-\,\tfrac{1}{4}\,\tfrac{1}{\sqrt{(N+1)(N+2)}}\,\dww\,\Big]\,,\label{B2}
\eea
where $N:=\w\c\dw$, and $\m$ is the continuous spin parameter.
We note that the operators $\mathbf{B}$, $\mathbf{B}_1$ and $\mathbf{B}_2$ are Hermitian (i.e. $\mathbf{B}^\dagger=\mathbf{B}$) with respect to the Hermitian conjugation rules
\be 
(\p_x^{\,\a})^\dag:=\,-\,\p_x^{\,\a} \,,\qquad (\dw^{\,\a})^\dag :=-\, \w^{\,\a} \,,\qquad (\w^{\,\a})^\dag:=-\,\dw^{\,\a} \,. \label{hermitian con}
\ee 
The action \eqref{Metsaev b} is invariant under the gauge transformation
\be 
\delta\, {{\Phi }} (x,\w)\,=\,\left(\w \cdot \p_x \,- \, \m \, \tfrac{1}{\sqrt{2(N+1)}} \,- \, \m ~ \w^2\,\tfrac{1}{\,{2(N+1)}\,\sqrt{2(N+2)}\,}\, \right) \chi(x,\w) \,,  \label{Gauge T m}
\ee 
where $\chi$ is the gauge transformation parameter introduced by the generating function
\be 
\chi(x,\omega)=\sum\limits_{s=1}^\infty\tfrac{1}{(s-1)!} \, \, \omega^{\m_{1}}   \ldots \omega^{\m_{s-1}} \, \,\chi_{\m_{1}\ldots\m_{s-1}}(x)\,.
\ee 

We note that this formulation of the CSP theory is constrained, that is, the gauge field $\Phi$ and the gauge transformation parameter $\chi$ are respectively double-traceless and traceless
\be 
(\,\dww\,)^2\,{\Phi}(x,\w)=0 \,, \qquad\qquad\qquad (\,\dww\,)\,\chi(x,\w)=0\,.
\ee 

By varying the action \eqref{Metsaev b} with respect to the gauge field $\Phi^\dagger$, we shall arrive at the bosonic CSP equation of motion
\be 
\big(\,\mathbf{B}+\mathbf{B}_1+\mathbf{B}_2\,\big)\,\Phi(x,\w)=0\,, \label{Eins}
\ee 
which after dropping a factor of $(\,1-\frac{1}{4}\,\w^2\,\dww\,)$ from its left-hand-side can be expressed as the following form
\begin{align}
&\bigg[ ~-\,\Box \,+\, \bigg(\omega \cdot \p_x\, -\,\m\,\tfrac{1}{\sqrt{2(N+1)}}\,-\,\m\,\omega^2\,\tfrac{1}{2(N+1)\,\sqrt{2(N+2)}}  \bigg) \nonumber \\
&~~~~~~\,\quad\,\times \le(\p_\omega \cdot \p_x
\,-\, \tfrac{1}{2}\,(\omega \cdot \p_x)\,\dw^2 \,-\,\m\,\tfrac{1}{\sqrt{2(N+1)}}  \,+\,\tfrac{1}{2}\,\m\,\tfrac{1}{\sqrt{2(N+2)}}\,\dw^2\,+\,\m\,\omega^2\,\tfrac{1}{4(N+2)\,\sqrt{2(N+3)}}\, \dw^2 \ri)\,\bigg]\,\Phi(x,\w)=0\,.\label{Ricc}
\end{align}
In comparison to the spin-two case, one can refer to \eqref{Eins} and \eqref{Ricc} as the Einstein-like and Ricci-like equations respectively. 
We note that in the helicity limit $\m\rightarrow0$, the above equation of motion reduces to a direct sum of all Fronsdal equations
\be 
\sum_{s=0}^{\infty}~\le[\,-\,\Box\,+\,\wdx\,\dwdx\,-\,\tfrac{1}{2}\,{\wdx}^2\,\dww~\ri]\,\phi_s(x,\w)=0\,,
\ee 
where $\phi_s$ was given by the generating function in \eqref{gen.}.

\subsection{Fermionic action}

Let us introduce the Dirac continuous spin gauge field by the generating function
\be 
 \Psi(x,\w) =\sum_{s=0}^{\infty} ~ \frac{1}{s!} \,\w^{\m_{1}}   \ldots \w^{\m_{s}}\, \Psi_{\m_{1}\ldots\m_{s}}(x)    \,,
\ee
where $\Psi_{\m_{1}\ldots\m_{s}}$ denote for all totally symmetric Dirac spinor-tensor fields of all half-integer spin $s+\frac{1}{2}$, and the spinor index is left implicit. The fermionic CSP action \cite{Metsaev:2017ytk}, in which the fermion field is a Dirac spinor, is then given by the Dirac continuous spin action
\be 
I_{_{CSP}}^{\,f}=\int d^4 x~\overline{\Psi}(x,\dw)~\big(\,\mathbf{F}+\mathbf{F}_1\,\big)~\Psi(x,\w)~\Big|_{\w=0}\,,\label{Metsaev f}
\ee 
where
\begin{align}
\mathbf{F}_{~}&:=\,\,i\,\,\Big[\, \ds-\ws\dwdx-\wdx\dws + \ws\ds\dws + \tfrac{1}{2}\,\ws\wdx\dww +\tfrac{1}{2}\,\w^2\dwdx\dws -\tfrac{1}{4}\,\w^2\ds \dww \Big]\,, \label{F}\\[5pt]
\mathbf{F}_1&:=\,\m\,\Big[\tfrac{1}{N+1}\,\le(1-\ws\dws-\tfrac{1}{4}\,\w^2\dww\ri)+\le(\ws-\tfrac{1}{2}\,\w^2\dws\ri)\tfrac{-\,i}{\sqrt{2(N+1)}}+\tfrac{-\,i}{\sqrt{2(N+1)}}\le(\dws-\tfrac{1}{2}\,\ws\dww\ri)\Big]  \label{F1}\,.
\end{align}
We note that operators $\mathbf{F}$, $\mathbf{F}_1$ are Hermitian (i.e. $\mathbf{F}^\dagger=\g^0\,\mathbf{F}\,\g^0$) with respect to the Hermitian conjugation rules \eqref{hermitian con}.

The action \eqref{Metsaev f} is invariant under the gauge transformation
\be 
\delta\,{\Psi}(x,\w)=\le(\,\w\cdot \p_x \,+\,\m\,\tfrac{1}{\sqrt{2(N+1)}} \,-\,i\,\m\,\ws\,\tfrac{1}{2(N+1)(N+2)} \,+\,\m\,\w^2\,\tfrac{1}{~[\,2(N+2)\,]^{3/2}} \,\ri){\tau}(x,\w)\,,
\label{Gauge T F 2}
\ee 
where $\tau$ is the spinor gauge transformation parameter introduced by the generating function
\be 
\tau(x,\omega)=\sum\limits_{s=1}^\infty\tfrac{1}{(s-1)!} \, \, \omega^{\m_{1}}   \ldots \omega^{\m_{s-1}} \, \,\tau_{\m_{1}\ldots\m_{s-1}}(x)\,.
\ee 
The formulation is constrained so as the spinor gauge field $\Psi$ and the spinor gauge transformation parameter $\tau$ are respectively triple gamma-traceless and gamma-traceless 
\be
\le(\,\dws\, \ri)^3 {{\Psi}}(x,\w) =0\,, \qquad \qquad \le(\,\dws \,\ri)\,\tau(x,\w)=0\,.
\ee 

By varying the action \eqref{Metsaev f} with respect to the gauge field $\overline{\Psi}$, one can easily obtain the Dirac CSP equation of motion
\be 
\big(\,\mathbf{F}+\mathbf{F}_1\,\big)\,\Psi(x,\w)=0\,,
\ee 
which after removing a factor of $(\,1-\frac{1}{2}\,\ws\,\dws-\frac{1}{4}\,\w^2\,\dww\,)$ from its left-hand-side will take the following form 
\begin{align}
&i\,\bigg[\,\ds-\,i \m ~\tfrac{1}{N+1} +\,\m \,\ws \, \tfrac{2}{~[2(N+1)]^{3/2}} - \Big(\omega \cdot \p_x +\,\m\,\tfrac{1}{~[2(N+1)]^{1/2}}+\,i\m\,\ws\, \tfrac{1}{2(N+1)(N+2)} +\,\m\,\w^2\,\tfrac{1}{~[2(N+2)]^{3/2}} \Big) \dws\bigg]\Psi(x,\w)=0\,.\label{Dirac CSP e}
\end{align}
We note that, similar to the previous section, in constrained formulation there are also two possibilities for presenting the fermionic CSP theory\,{\color{blue}\footnote{Notice that, in contrast to the unconstrained formulation, here there exists just one possibility for expressing the fermionic HS theory, which is the Fang-Fronsdal formalism \cite{Fang:1978wz}.}}. One possibility is the one we have stated in above, however there exists another possibility which obtains by converting $i \rightarrow -\,i$ in relations of \eqref{F1}, \eqref{Gauge T F 2}, \eqref{Dirac CSP e} that has been applied in \cite{Najafizadeh:2017acd} (see also appendix \ref{I op}).

\vspace{.3cm}

\noindent\textbf{Remark:} 

\vspace{.2cm}

\noindent Let us here pursue again the discussion we had in the previous section about Weyl equations. Referring to the issue and using the notation we applied in that section, one can write the Dirac CSP equation of motion \eqref{Dirac CSP e} as 
\be 
\le( \begin{array}{cc}
	\mathcal{M} & i\,{\Sigma}+i\,{\Xi} \\
	i\,{\overline\Sigma}+i\,\overline{\Xi} &  {\mathcal{\overline M}}\\
\end{array} \ri) \le( \begin{array}{c}
	\Psi_L  \\
	\Psi_R \\
\end{array} \ri)=0\,,\label{Weyl CSP 2}
\ee 
where 
\begin{align}
{\mathcal{ M}}&:=\m\,\tfrac{1}{N+1}\,+\,\m\,\tfrac{1}{2N(N+1)}\,(\s\c\w)(\bar\s\c\dw)\,,\\[5pt]
\Sigma&:=\s\c\p - \wdx(\s\c\dw) \,,\\[5pt]
\Xi&:=\m\,(\s\c\w)\,\tfrac{2}{~[2(N+1)]^{3/2}}\,-\,\m\,\tfrac{1}{~[2(N+1)]^{1/2}}\,(\s\c\dw)\,
-\,\m\,\w^2\,\tfrac{1}{~[2(N+2)]^{3/2}}\,(\s\c\dw)\,.
\end{align} 
It is clear, when $\m\neq 0$, the operator ${\mathcal{ M}}$ is non-zero and as a result the equation \eqref{Weyl CSP 2} does not decompose into two Weyl equations. However, in the helicity limit $\m\rightarrow0$, which the higher spin equations should be reproduced, the operators $\mathcal{M}$ and $\Xi$ vanish, and consequently the equation \eqref{Weyl CSP 2} decouples into two Weyl higher spin equations:
\be 
i\,{\overline\Sigma}\,\Psi_L=0\,,\qquad \qquad\qquad i\,{\Sigma}\,\Psi_R=0\,.
\ee 
Therefore, one may conclude that the continuous spin parameter $\m$ (which has a dimension of mass) in the Dirac CSP equation plays a role as mass in the massive Dirac spin-$\frac{1}{2}$ equation. Accordingly, one can observe that although continuous spin particle is a massless object, there would be no Weyl continuous spin equation, at least in its two formulations which we have studied in this paper. We recall that the existence of Weyl equations was dependent on formulation we use, so as at $\m=0$ there was no weyl equations based on unconstrained formulation, while there exists in constrained one.


\subsection{Supersymmetry transformations}

In previous subsections, we discussed constrained formulation of the bosonic and fermionic continuous spin gauge field theories in 4-dimensional flat space-time. At this stage we are ready to provide supersymmetry transformations for the $\mathcal{N}=1$ continuous spin supermultiplet, in which CSP and CSPino are respectively a complex scalar and a Dirac continuous spin fields. We acquire that the supersymmetry continuous spin action (sum of the bosonic \eqref{Metsaev b} and fermionic \eqref{Metsaev f} continuous spin actions)  
\be 
I_{_{CSP}}^{^{\,SUSY}}=I_{_{CSP}}^{\,b}+I_{_{CSP}}^{\,f} \label{susy action met}
\ee 
is invariant under the following supersymmetry transformations
\begin{tcolorbox}[ams align, colback=white!98!black]
\hspace{-.25cm}&\d \,\Phi(x,\w) ={\scriptstyle\sqrt{2}}\,\bar\epsilon~\big(\tfrac{\,1\,+\,\g^5\,}{2}\big)~{\Psi}(x,\w)+i\,\bar\epsilon\,\,\ws\,\tfrac{1}{\sqrt{(N+1)}}~\big(\tfrac{\,1\,-\,\g^5\,}{2}\big)~{\Psi}(x,\w)\,, \label{12c}\\[6pt]
\hspace{-.25cm}&\d \,\Psi(x,\w) =\bigg{\{}-\, \tfrac{i}{\sqrt{2}}\le[\,2\,\ds - \ws\,\tfrac{1}{(N+1)}\,(\dw\c\p_x)+\ws\,\ds\,\tfrac{1}{(N+1)}\,\dws - \ws\,(\w\c\p_x)\,\tfrac{1}{2(N+2)}\,\dww\,\ri] \label{34c}\\[6pt]
&~~~~~~~~~~~\qquad\,+\,\tfrac{1}{\sqrt{(N+1)}}\,\le[\,\ds\,\dws- \dwdx-\tfrac{1}{2}\,\wdx\,\dww\,\ri]\,+\nonumber\\[6pt]
&\hspace{-.45cm}\tfrac{1}{\sqrt{2}}\,\m\,\Big[~\tfrac{1}{N+1}-\w^2\,\tfrac{1}{4(N+2)(N+3)}\,\dww-\,\tfrac{1}{\sqrt{4(N+1)(N+2)}}\,\dww-i\,\ws\,\tfrac{1}{\sqrt{2}(N+1)^{3/2}}+i\,\ws\,\tfrac{1}{2(N+1)\sqrt{2(N+2)}}\,\dww~\Big]\bigg{\}}\big(\tfrac{\,1\,-\,\g^5\,}{2}\big)\,\epsilon~\Phi(x,\w),\nonumber
\end{tcolorbox}
\hspace{-.5cm}
where the supersymmetry transformation parameter $\epsilon$ is a Dirac spinor object, and $\Phi$, $\Psi$ are respectively the complex scalar and Dirac CSP fields. Using the above transformations, it is tedious but straightforward to check the closure of the SUSY algebra. We find that the algebra closes on-shell 
\begin{align}
[\,\d_1\,,\,\d_2\,]\,\Phi(x,\w)&=-\,2\,i\,(\bar\epsilon_2\,\ds\,\epsilon_1)\,\Phi(x,\w)\,,\label{susy al 1}\\[5pt] 
[\,\d_1\,,\,\d_2\,]\,\Psi(x,\w)&\approx-\,2\,i\,(\bar\epsilon_2\,\ds\,\epsilon_1)\,\Psi(x,\w)\,+\,\hbox{gauge transformation}\,,\label{susy al 2}
\end{align}
up to a gauge transformation which is proportional to \eqref{Gauge T F 2}. 

\vspace{.5cm}

\noindent\textbf{Remarks:} Most of remarks in the preceding section are valid here, however, let us add a few points concerning the supersymmetry transformations \eqref{12c} and \eqref{34c}: 

\vspace{.2cm}


\begin{itemize}

\item The gamma fifth matrix $\g^5$ is responsible for closure of the SUSY algebra, so as by dropping the $\g^5$ from the above supersymmetry transformations, the SUSY action \eqref{susy action met} will remain still invariant under \eqref{12c} and \eqref{34c}.

\item It is notable to see that the SUSY CSP variation of boson field \eqref{12c} contains two terms. The first term is proportional to the SUSY variation of the half-integer spin supermultiplet \eqref{hss1c1}, and the second term is corresponding to the SUSY variation of the integer spin supermultiplet \eqref{hss1c}.

\item Moreover, one can observe that the first line in the SUSY CSP variation of fermion field \eqref{34c} is identical to the SUSY variation of the half-integer spin supermultiplet \eqref{hss2c1}, while the second line in \eqref{34c} is proportional to the integer spin supermultiplet \eqref{hss2c}.


\end{itemize}

\subsection{Helicity limit} \label{hl}

Let us go on the discussion was carried out about the helicity limit in the previous section. However, here, the formulation is constrained and one expects to reach to the well-known result of \cite{Curtright:1979uz} in the helicity limit. To be more precise, in the helicity limit, result of the supersymmetric higher spin theory, i.e. supersymmetry transformations of half-integer and integer spin supermultiplets \eqref{hss1c1},\eqref{hss2c1} and \eqref{hss1c},\eqref{hss2c} discussed in the section \ref{HS review}, are expected to be recovered. However, we note that the chiral supermultiplet $(\,{\scriptstyle 0}\,,\, {\scriptstyle 1/2}\,)$ was irrelevant for higher spins and was not discussed, while here in the helicity limit of the continuous spin theory it may be reproduced. In order to make clear the discussion, let us take into account the CSP supermultiplet \eqref{CSP super}, in which the complex scalar CSP and Dirac CSP fields are given respectively by \eqref{phi} and \eqref{psi}:
\be 
\bigg(~~\Phi(x,\w)~~,~~\Psi(x,\w)~~\bigg) \qquad \Longleftrightarrow \qquad
\bigg(~~\sum_{s=0}^{\infty} ~ \tfrac{1}{s!} ~\w^{\m_{1}}   \ldots \w^{\m_{s}}\, \Phi_{\m_{1}\ldots\m_{s}}(x)~~,~~\sum_{s=0}^{\infty} ~ \tfrac{1}{s!} ~\w^{\m_{1}}   \ldots \w^{\m_{s}}\, \Psi_{\m_{1}\ldots\m_{s}}(x)~~\bigg)\,.\label{CSP super 2}
\ee
On the other side, in the helicity limit $\m\rightarrow0$, we know that the continuous spin representation becomes reducible and decomposes into the direct sum of all helicity representations. Therefore, at $\m=0$, one can expect that the above supermultiplet decomposes into a direct sum of the chiral supermultiplet as well as the well-known half-integer and integer spin supermultiplets of the higher spin theory, i.e. 
\be 
\Big(~0~,~\tfrac{1}{2}~\Big)~~\oplus~~ \sum_{s=1}^{\infty}~\Big(~s~,~s+\tfrac{1}{2}~\Big)
~~\oplus~~ \sum_{s=0}^{\infty}~\Big(~s+\tfrac{1}{2}~,~s+1~\Big)\,.
\ee 
In what follows, let us discuss and attempt to reproduce each case separately.
\vspace{.3cm}

\noindent\textbf{Chiral supermultiplet\,$(\,{\scriptstyle 0}\,,\, {\scriptstyle 1/2}\,)$\,:} 

\vspace{.1cm}

\noindent At $\m=0$, if one just considers spin-$0$ and spin-$\frac{1}{2}$ fields in the infinite towers of spins \eqref{CSP super 2}, there would be no $\w$ in the gauge fields and consequently the act of $\w$-dependent derivatives on the gauge fields vanish. Thus, the supersymmetry transformations \eqref{12c} and \eqref{34c} reduce to those for the chiral supermultiplet (appendix \ref{Wess-Zu})
\begin{align}
\d \,\phi(x) &={\scriptstyle\sqrt{2}}~\overline{\epsilon_R}~{\psi}_L(x)\,,\label{chiral 1}\\
\d \,\psi_L(x) &=-\,i\,{\scriptstyle\sqrt{2}}~ \ds~\phi(x)~\epsilon_R\,,\label{chiral 2}
\end{align}
where we have considered decomposition of the Dirac field $\psi=\psi_L +\psi_R$ and the supersymmetry transformation parameter $\epsilon=\epsilon_L + \epsilon_R$ in terms of Weyl spinors, and took into account
\be
\psi_L= \big(\tfrac{\,1\,+\,\g^5\,}{2}\big)\,\psi=L\,\psi\,,\qquad \qquad
\psi_R= \big(\tfrac{\,1\,-\,\g^5\,}{2}\big)\,\psi=R\,\psi\,.
\ee 
\vspace{.3cm}

\noindent\textbf{Half-integer\,$(\,s\,,\, s\, {\scriptstyle +\, 1/2}\,)$\, and integer\,$(\, s\, {\scriptstyle +\, 1/2}\,,\,s\,{\scriptstyle +\,1}\,)$\, spin supermultiplets\,:} 

\vspace{.1cm}

\noindent Ignoring the chiral supermultiplet, one can redefine the supersymmetry transformation parameter
\be 
\varepsilon :=\big(\tfrac{\,1\,-\,\g^5\,}{2}\big)\,\epsilon\,,
\ee 
and as a result, at $\m=0$, one can illustrate that \eqref{12c} and \eqref{34c} are a direct sum of the reducible supersymmetry transformations of the reducible half-integer spin supermultiplet $(\,s\,,\, s\, {\scriptstyle +\, 1/2}\,)$
\bea
\d \,\phi(x,\w) &=&{\scriptstyle\sqrt{2}}\,\bar\varepsilon~{\psi}(x,\w)\,, \label{12ca}\\[3pt]
\d \,\psi(x,\w) &=&-\, \tfrac{i}{\sqrt{2}}\le[\,2\,\ds - \ws\,\tfrac{1}{(N+1)}\,(\dw\c\p_x)+\ws\,\ds\,\tfrac{1}{(N+1)}\,\dws - \ws\,(\w\c\p_x)\,\tfrac{1}{2(N+2)}\,\dww\,\ri]\,\varepsilon~\phi(x,\w)\,, \label{34ca}
\eea
as well as the reducible supersymmetry transformations of the reducible integer spin supermultiplet {$(\, s\, {\scriptstyle +\, 1/2}\,,\,s\,{\scriptstyle +\,1}\,)$}
\bea
\hspace{-4.5cm}\,~~\d \,\phi(x,\w) &=&i\,\bar\varepsilon\,\,\ws\,\tfrac{1}{\sqrt{(N+1)}}~{\psi}(x,\w)\,, \label{12cb}  \\[0pt]
\hspace{-4.5cm}~\,~\d \,\psi(x,\w) &=&\tfrac{1}{\sqrt{(N+1)}}\,\le[\,\ds\,\dws- \dwdx-\tfrac{1}{2}\,\wdx\,\dww\,\ri]\,\varepsilon~\phi(x,\w)\,.
\eea
In these two above SUSY transformations, the bosonic field $\phi(x,\w)$ is a complex higher spin field, and the fermionic field $\psi(x,\w)$ is a Dirac higher spin field. However, as we know, such transformations are reducible and can reduce, respectively, to the well-known irreducible supersymmetry transformations of the irreducible half-integer spin supermultiplet \eqref{hss1c1},\eqref{hss2c1} and integer spin supermultiplet \eqref{hss1c},\eqref{hss2c}, which were presented by Curtright in \cite{Curtright:1979uz}.


Let us close this section with a remark on the helicity limit where we try to demonstrate that in the limit $\mu=0$, we get the correct reducible supersymmetry transformations of the reducible higher spin supermultiplets. In the SUSY CSP transformations \eqref{12c}, \eqref{34c}, the fermionic CSP field is a Dirac field, however, at $\m=0$, we arrive at \eqref{chiral 1} and \eqref{chiral 2}, illustrating that the chiral supermultiplet $(\,{\scriptstyle 0}\,,\, {\scriptstyle 1/2}\,)$ involves only the left-handed part of the Dirac spin-$\tfrac{1}{2}$ field, then the natural question is what happens with the right-handed part of the Dirac spin-$\tfrac{1}{2}$ field? It seems what happens is that this right-part combines with a real (or imaginary) part of the spin-$1$ field into the integer spin-$1$ supermultiplet $(\,{\scriptstyle 1/2}\,,\, {\scriptstyle 1}\,)$, while the imaginary (or real) part of the spin-$1$ field couples to a left-haded (or right-handed) Weyl part of the spin-$\tfrac{3}{2}$ Dirac field which thus form the half-integer spin-$\tfrac{3}{2}$ supermultiplet $(\,{\scriptstyle 1}\,,\, {\scriptstyle 3/2}\,)$ and so on and so forth\,{\color{blue}\footnote{We acknowledge Dmitri Sorokin for bringing our attention to this question and his clarifications.}}.



\section{Relationship of supersymmetry transformations} \label{sec. rel}

In this section we aim to make relationship between our unconstrained continuous spin SUSY transformations \eqref{hss1}, \eqref{hss2} and the constrained SUSY transformations \eqref{12c}, \eqref{34c}. For this purpose, we begin from the unconstrained CSP SUSY transformations, and will pursue three steps; performing the Fourier transformation, applying field redefinition, and changing
of auxiliary space variable. We note that, however, one can follow a reverse approach by starting from the constrained CSP SUSY transformations.

\subsection{Fourier transformation}
Let us multiply the SUSY variation of boson field \eqref{hss1} by $\d'(\e^2+1)$, and the SUSY variation of fermion field \eqref{hss2} by $\d'(\e^2+1)(\es-i\,)$ to the left which become
\begin{align}
\hspace{-.28cm}\delta\,\Big[\d'(\e^2+1)\,\Phi(x,\e)\Big]&= \,\tfrac{1}{\sqrt{2}}\,\, \bar\epsilon\,\,\big(\,1+\g^5\,\big)\,\d'(\e^2+1)\,\big(\es-i\,\big)\,\Psi(x,\e)\,,\label{fou1}\\[4pt]
\hspace{-.28cm}\delta\,\Big[\d'(\e^2+1)(\es-i\,)\,\Psi(x,\e)\Big]&=-\,\tfrac{1}{\sqrt{2}}\, \le[\,\ds\,(\es+i\,)-\e\c\p_x+\tfrac{1}{2}\,(\de\c\p_x+\m)\,(\e^2+1)\,\ri]\big(\,1-\g^5\,\big)\,\d'(\e^2+1)\,\Phi(x,\e)~\epsilon.\label{fou}
\end{align}
We then perform a Fourier transformation in the auxiliary space variable $\e^\m$ to express relations \eqref{fou1}, \eqref{fou} in their Fourier-transformed auxiliary space, i.e. $\w$-space, via
\bea
\widetilde{\Phi}(x,\w)&\equiv& \int d^4\eta \,\,e^{-\,i\eta\cdot\omega}~
\delta'(\eta^2 +1)\,\Phi(x,\e),  \label{1}
\\
\widetilde{\Psi}(x,\w)&\equiv& \int d^4\eta \,\,e^{-\,i\eta\cdot\omega}~
\delta'(\eta^2 +1) (\es - i\,)\,\Psi(x,\e)\,. \label{2} 
\eea 
Notice that the fields in the left-hand-sides of the latter are constrained while the ones in the right-hand-sides are unconstrained. More precisely, the equations \eqref{1} and \eqref{2} can be understood respectively as the general solutions of the double traceless-like and the triple gamma-traceless-like conditions
\begin{align}
(\dww -1\,)^2\,\widetilde{\Phi}(x,\w)&=0\,,\label{trace-like1}\\
(\dws-1\,)(\dww-1\,)\,\widetilde{\Psi}(x,\omega)&=0\,.\label{trace-like2}
\end{align}
Using \eqref{1}, \eqref{2}, we perform the Fourier transformation over the auxiliary variable $\e$, and rewrite \eqref{fou1} and \eqref{fou} in $\w$-space, which become
\begin{align}
\d\,\widetilde{\Phi}(x,\w)&= \tfrac{1}{\sqrt{2}}\,\bar\epsilon\,\,\big(\,1+\g^5\,\big)\,\widetilde{\Psi}(x,\w)\,,\label{fou22}\\
\d\,\widetilde{\Psi}(x,\w)&=i\,\tfrac{1}{\sqrt{2}}\, \le[\,\ds\,(\dws-1\,)-(\dw\c\p_x)-\tfrac{1}{2}\,(\w\c\p_x+i\,\m\,)\,(\dww-1\,)\,\ri]\,\widetilde{\Phi}(x,\e)~\big(\,1-\g^5\,\big)\,\epsilon\,.\label{fou2}
\end{align}

\subsection{Field redefinition}

As it can be seen from \eqref{trace-like1}, \eqref{trace-like2}, the gauge fields $\widetilde{\Phi}$ and $\widetilde{\Psi}$ in \eqref{fou22}, \eqref{fou2} are respectively double traceless-like and triple gamma-traceless-like. However, one can apply a field redefinition to rearrange traces. We have elaborated such rearrangement in detail in \cite{Najafizadeh:2017acd}. In fact, by applying the following fields redefinition
\begin{align}
\widetilde{\Phi} (x,\omega)&=\mathbf{P}_\Phi~\Phi (x,\omega)\,, \quad \quad\quad
~\mathbf{P}_\Phi:= \sum_{k=0}^{\infty}~\omega^{\,2k}~ \frac{1}{~2^{\,2k} ~ k!~ (N+1)_k~}\,, 
\label{P_phi text}\\
\widetilde{\Psi} (x,\omega)&=\mathbf{P}_\Psi~\Psi(x,\omega)\,, \quad \quad\quad\,
\mathbf{P}_\Psi :=\sum_{k=0}^{\infty}~\le[(\ws)^{2k} + 2k(\ws)^{2k-1}\ri]
\frac{1}{~2^{\,2k}~k!~(N+1)_k~} 
\label{P say}\,,
\end{align}
where $N:={\w\c\dw}$ and $(a)_k$ is the rising Pochhammer symbol \eqref{Pochhammer}, one can convert the double traceless-like condition \eqref{trace-like1} to the double traceless one $(\dww)^2\,\Phi(x,\w)=0$, and reduce the triple gamma-traceless-like condition \eqref{trace-like2} to the triple gamma-traceless one $(\dws)^3\,\Psi(x,\w)=0$. In terms of these redefined CSP fields $\Phi(x,\w)$ and $\Psi(x,\w)$, one can rewrite relations \eqref{fou22}, \eqref{fou2} as 
\begin{align}
\d \,\Phi(x,\w) 
&=\tfrac{1}{\sqrt{2}}\,\bar\epsilon\,\,\big(\,1+\g^5\,\big)\,\le(1+\ws\,\tfrac{1}{2(N+1)}\ri)\,{\Psi}(x,\w)\,,\label{12cc}\\[6pt]
\d \,\Psi(x,\w) &=-\, \tfrac{i}{\sqrt{2}}\,\bigg[~\ds - \ws\,\tfrac{1}{2(N+1)}\,(\dw\c\p_x)+\ws\,\ds\,\tfrac{1}{2(N+1)}\,\dws - \ws\,(\w\c\p_x)\,\tfrac{1}{4(N+2)}\,\dww \label{34cc}\\[6pt]
&\qquad\quad\qquad\,+ \dwdx-\ds\,\dws+\tfrac{1}{2}\,\wdx\,\dww\nonumber\\[6pt]
&\,~~+\,\tfrac{1}{2}\,i\,\m\,\Big(~\tfrac{1}{N+1}+\dww-\ws\,\tfrac{1}{2(N+1)^2}-\ws\,\tfrac{1}{2(N+1)}\,\dww-\w^2\,\tfrac{1}{4(N+2)(N+3)}\,\dww~\Big)~+~\mathcal{O}(\w^3)\,\bigg]\,\Phi(x,\w)\,(\,1-\g^5\,) \,\epsilon\,.\nonumber
\end{align}
To obtain the relation \eqref{12cc}, one can simply use \eqref{P_phi text}, \eqref{P say} in \eqref{fou22} and apply the relation between operators $\mathbf{P}_\Phi$ and $\mathbf{P}_\Psi$, given in \eqref{psiphi}. This in turn leads to \eqref{12cc} by removing the operator $\mathbf{P}_\Phi$ in both sides of \eqref{fou22} from the left. In addition, the relation \eqref{34cc}, straightforwardly, can be acquired by plugging \eqref{P_phi text} and \eqref{P say} in \eqref{fou2}, then multiplying the obtained relation by the inverse of $\mathbf{P}_\Psi$ \eqref{ps-1} to the left, and finally applying relations \eqref{11}-\eqref{33}. We note that, in \eqref{34cc}, appeared terms of order $\w^3$ eliminate in variation of the SUSY action due to the constraint $\overline{\Psi}(x,\dw)\,(\ws)^3=0$, so we do not consider such terms in rest of this section.

\subsection{Change of variable}

As final step, let us make a change of variable in the auxiliary space $\w$ by shifting   
\be 
\w^\a\, \,\longrightarrow \,\,i\,\w^\a\,{\scriptstyle\sqrt{2(N+1)}}\, ,
\ee 
which in turn leads to the following changes
\be 
\dw^{\,\a} \,\,\longrightarrow \,\,-\,\tfrac{i}{\sqrt{2(N+1)}}\,\, \dw^{\,\a}\,,
\qquad \w^2 \, \,\longrightarrow \,\, -\,\w^2\,{\scriptstyle\sqrt{4(N+1)(N+2)}}\,,
\qquad \dww \, \,\longrightarrow \,\, -\,\tfrac{1}{\sqrt{4(N+1)(N+2)}}\,\dww\,.
\ee 
If one applies these changes in relations \eqref{12cc} and \eqref{34cc}, ones convert to
\begin{align}
\d \,\Phi(x,\w) 
&=\tfrac{1}{\sqrt{2}}\,\bar\epsilon\,\,\big(\,1+\g^5\,\big)\,\Big(\,1+i\,\ws\,\tfrac{1}{\sqrt{2(N+1)}}\,\Big)\,{\Psi}(x,\w)\,,\\[6pt]
\d \,\Psi(x,\w) &=-\, \tfrac{i}{\sqrt{2}}\,\bigg[~\ds - \ws\,\tfrac{1}{2(N+1)}\,(\dw\c\p_x)+\ws\,\ds\,\tfrac{1}{2(N+1)}\,\dws - \ws\,(\w\c\p_x)\,\tfrac{1}{4(N+2)}\,\dww \label{34ccc}\\[6pt]
&\qquad\quad\qquad\,-\,\tfrac{i}{\sqrt{2(N+1)}}\le( \dw\c\p_x-\ds\,\dws+\tfrac{1}{2}\,\wdx\,\dww\ri)\nonumber\\[6pt]
&\hspace{-1.5cm}+\tfrac{i\,\m}{2}\,\Big(\,\tfrac{1}{N+1}-\,\tfrac{1}{\sqrt{4(N+1)(N+2)}}\,\dww-i\,\ws\,\tfrac{1}{\sqrt{2}(N+1)^{3/2}}+i\,\ws\,\tfrac{1}{2(N+1)\sqrt{2(N+2)}}\,\dww-\w^2\,\tfrac{1}{4(N+2)(N+3)}\,\dww\,\Big)\bigg]\,\Phi(x,\w)\,(1-\g^5)\,\epsilon\,.\nonumber
\end{align}
These supersymmetry transformations are precisely the ones we presented in \eqref{12c}, \eqref{34c} for constrained formulation of the CSP theory. Therefore, by following the above three steps, we could make a precise relation between two separate set of SUSY transformations, unconstrained \eqref{hss1}, \eqref{hss2} and constrained \eqref{12c}, \eqref{34c} ones.



%

\section{Conclusions and outlook}\label{conclu}

In this paper, we first reviewed the supersymmetric higher spin gauge theory and obtained supersymmetry transformations for the $\mathcal{N}=1$ half-integer and integer spin supermultiplets, studied long time ago by Curtright \cite{Curtright:1979uz}. Nevertheless, our review was based on the generating functions and we dealt with operators facilitating calculations. In addition, the review included a method in detail which we applied to find the SUSY CSP transformations.

\vspace{.2cm}

Then, taking into account the Schuster-Toro action \cite{Schuster:2014hca} and its fermionic analogue \cite{Najafizadeh:2015uxa}, we supersymmetrized unconstrained formulation of the continuous spin gauge field theory. To this end, we provided supersymmetry transformations \eqref{hss1}, \eqref{hss2} for the $\mathcal{N}=1$ supermultiplet which leave the SUSY continuous spin action \eqref{48} invariant. Since a CSP (bosonic or fermionic) has infinite physical degrees of freedom per space-time point, we observed that the number of real CSP fields should be equal in the $\mathcal{N}=1$ CSP supermultiplet (which may be held for $\mathcal{N}>1$). Therefore, we took into account a CSP supermultiplet \eqref{cspinp}, in which CSP is a complex scalar continuous spin field and CSPino is a Dirac continuous spin field. We note that in the frame-like approach, authors of \cite{Buchbinder:2019kuh} provided supersymmetry transformations for the $\mathcal{N}=1$ infinite spin supermultiplet containing four real fields; a pair of massless bosonic CSP fields with opposite parity, and a pair of massless fermionic CSP fields. Therefore, in this regard, the number of real fields we used to supersymmetrize the CSP theory in the metric-like approach is compatible with \cite{Buchbinder:2019kuh}.

\vspace{.2cm}

We then took the helicity limit of the CSP theory, and supersymmetrized unconstrained formulation of the higher spin gauge theory \`a la Segal, given by the bosonic \cite{Segal:2001qq} and fermionic \cite{Najafizadeh:2018cpu} actions, in 4-dimensional flat space-time. To supersymmetrize this theory, similar to the CSP case, we considered the $\mathcal{N}=1$ higher spin supermultiplet in which HS is a complex higher spin field and the so-called ``HSpino'' is a Dirac higher spin field. In both cases, continuous spin and higher spin, the fact that we should have a complex field in the supermultiplet is related to the presence of the spin-$0$ field in the spectrum, which should be complex in the chiral supermultiplet\,{\color{blue}\footnote{We thank Dmitri Sorokin for comments and pointing out this issue.}}. We recall that in the supersymmetric higher spin theory \`a la Fronsdal \cite{Curtright:1979uz}, the spin-$0$ field does not exist in the spectrum while here there exists.

\vspace{.2cm}

Afterwards, building on the Metsaev actions in 4-dimensional flat space-time \cite{Metsaev:2016lhs}, \cite{Metsaev:2017ytk}, we supersymmetrized constrained formulation of the continuous spin gauge theory by providing supersymmetry transformations \eqref{12c}, \eqref{34c}. In both formulations, the gamma fifth was employed to close the algebra and we illustrated that the SUSY algebra closes on-shell up to a gauge transformation. Moreover, we demonstrated that although CSP is a massless elementary particle, the continuous spin parameter $\m$ in the theory plays a role of mass, and thus the Dirac continuous spin equation can not be decoupled into Weyl equations. We also made a relationship between two set of unconstrained and constrained supersymmetry transformations by performing a Fourier transformation, field redefinition and change of variable.  

\vspace{.2cm}
As we know, in the helicity limit $\m\rightarrow0$, the continuous spin representation becomes reducible and decomposes into the direct sum of all helicity representations. Therefore, in the limit, the bosonic (fermionic) CSP field gives rise to an infinite set of the bosonic (fermionic) higher spin fields in which each spin appears only once. Let us mention that there is a different formulation in which similar infinite sets of higher spin fields appear. In this formulation the infinite sets of higher spin fields are described as scalar and spinor fields in the so-called tensorial (or hyper) spaces (for a review and references see \cite{Sorokin:2017irs}). Supersymmetric higher spin models constructed in hyperspace \cite{Bandos:1999qf,Bandos:2004nn,Florakis:2014kfa,Florakis:2014aaa} describe infinite-dimensional higher spin supermultiplets and thus differ from the conventional higher spin supermultiplets obtained in the helicity limit of the supersymmetric CSPs.

\vspace{.2cm}
Constrained formulation of the continuous spin theory \`a la Fronsdal is more favorable for higher spin community, however, it seems calculations in the unconstrained formulation \`a la Segal or Schuster-Toro are more convenient. For instance, at a glance, one can see that the form of supersymmetry transformations in \eqref{hss1}, \eqref{hss2} are more brief in comparison with those in \eqref{12c}, \eqref{34c}, however both are equivalent and can be converted to each other. Therefore, it would be interesting to establish the massive higher spin gauge theory in unconstrained formulation and find its supersymmetry transformations which will probably take a simple form but equivalent to existing shapes \cite{Zinoviev:2007js}. In addition, it would be nice to construct a supersymmetric massive higher spin gauge theory in constrained formulation, such that by taking the continuous spin limit ($m\rightarrow 0$\,, $s \rightarrow \infty$\, while $ms=\m=\mbox{constant}$) converts to the result of this paper. It is also interesting to develop cubic interaction vertices for the $\mathcal{N}=1$ arbitrary spin massless supermultiplets \cite{Metsaev:2019dqt}, \cite{Metsaev:2019aig} to the continuous spin gauge theory.




\vspace{.5cm}

\noindent\textbf{\centerline{Acknowledgements:}} 

\vspace{.3cm}

\noindent We are extremely grateful to Dmitri Sorokin for helpful discussions, comments and patiently answering many questions. We would like to thank Hamid Reza Afshar, Xavier Bekaert, Thomas Curtright, Mohammad Khorrami, Ruslan Metsaev, Antoine Van Proeyen, Mohammad Mahdi Sheikh-Jabbari and Yuri Zinoviev for useful discussions. %
%
%
We also appreciate Joseph Buchbinder and Dario Francia for comments. We thank the Erwin Schr\"odinger International Institute for Mathematics and Physics (ESI) for support and hospitality during the Workshop ``\,\href{http://quark.itp.tuwien.ac.at/~grumil/ESI2019/}{\color{black}Higher spins and holography}\,'' in Vienna (March 11 - April 5, 2019) where some part of this work was carried out, as well as for holding the lecture course of ``\,\href{https://www.esi.ac.at/events/e286/}{\color{black}Supergravity}\,'' by Antoine Van Proeyen (March 7 - 29, 2019) which was fruitful for this work. We also acknowledge Daniel Grumiller and Mohammad Mahdi Sheikh-Jabbari for financial support during this visit via the Iran-Austria IMPULSE project grant supported and run by Kharazmi University and OeAD-GmbH.


\appendix

\section{Conventions}\label{conven}

We use the mostly minus signature for the metric and work in the 4-dimensional Minkowski space-time. The convention
\be
\p_{\,\w_{\,\n}}=\frac{\p}{\p\,{\w^{\,\n}}}\,, \quad\quad\quad   \n=0,1,2,3\,,
\ee
and the following commutation relations
\be
\le[\,\p_{\omega }^{\,\a} \,,\, \omega^{\,\b}\,\ri] =  \eta^{\,\a\b}\,,  \label{commutation 1}
\quad \quad \quad \quad
\le[\,{\dww} \,,\, \omega^{\,2}\,\ri] =  4\, (N + 2) \,, \quad  N:= \omega \cdot \p_\omega\,,
\ee
are used\,. The hermitian conjugates in $\w$-space and $\e$-space introduce as
\be
(\p_x^{\,\a})^\dag:=\,-\,\p_x^{\,\a} \,,\qquad (\dw^{\,\a})^\dag :=-\, \w^{\,\a} \,,\qquad (\w^{\,\a})^\dag:=-\,\dw^{\,\a} \,,\qquad (\de^{\,\a})^\dag := -\,\de^{\,\a}\,,\qquad (\e^{\,\a})^\dag := \e^{\,\a} \,.
\label{hermitian conjugates 2}
\ee
%
For the 4-dimensional Dirac gamma-matrices we use
the conventions
\be
\big\{\,\g^\a \,,\,\g^\b \,\big\}=2\,\e^{\,\a\b}
\,,
\quad\quad\quad\quad
(\,\g^\a\,)^{\,\dag} = \g^0\,\g^\a\,\g^0 \,,
\quad\quad\quad\quad
(\,\g^0\,)^{\,\dag} = \g^0\,,
\quad\quad\quad\quad
(\,\g^0\,)^{\,2} = 1\,,
\label{anti comm gamma}
\ee
\be 
\ds:=\g^\m\,\p_\m\,,\quad\quad\quad \dws:=\g^\m\,{\dw}_\m\,,\quad\quad\quad\ws:=\g^\m\,\w_\m\,,
\quad\quad\quad\es:=\g^\m\,\e_\m\,,\quad\quad\quad\overline\Psi=\Psi^\dagger\,\g^0\,,
\ee 
\be
\big\{\,\dws \,,\, \ws \,\big\} =2\,(N + {2})\,,\quad\quad\quad\g^5=i\,\g^0\,\g^1\,\g^2\,\g^3\,,\quad\quad\quad (\,\g^5\,)^{\,2}=1\,,
\quad\quad\quad \big\{\,\gamma^\a \,,\, \gamma^5 \,\big\}=0 
\,.
\ee

\section{Proof}\label{proof}

Considering the Schuster-Toro action \eqref{S-T action} and its fermionic analogue \eqref{Mojtaba}, we aim to prove that a $\d \Psi$ can not be found if we start with the ansatz $\d\Phi=\a\, \bar\epsilon \, \Psi$ which $\a$ is an arbitrary parameter. To this end, let us consider the ansatz as
\be 
\d\Phi=\a\, \bar\epsilon \, \Psi \,, \qquad\qquad \d\Psi=\a\,X\,\Phi\,\epsilon\,, \label{ansatz}
\ee 
where $\Phi$ and $\Psi$ are considered to be, respectively, a real scalar CSP field and a Majorana CSP field, and $X$ is an operator which we aim to find (if any). It is convenient to find that the invariance of a sum of the real scalar CSP action and the Majorana CSP action under the ansatz \eqref{ansatz} leads to the following relation (for simplicity we set $\m=0$ without losing any accuracy of the proof)
\be 
-\,\frac{1}{2}\,\le[-\,2\,\p_x\c\p_x+2\,(\e\c\p_x)(\de\c\p_x)-(\e^2+1)(\de\c\p_x)^2\,\ri]= \le[\,(\es+i\,)\,\ds - (\e^2+1)(\de\c\p_x)\,\ri]\,X\,.\label{eq}
\ee  
It is seen that the left-hand-side of \eqref{eq} is quadratic in $\p_x$, and as no function of $x$ is involved in the left-hand-side, $X$ should be linear in $\p_x$. Thus, one can consider
\be 
X=-\,\tfrac{1}{2}\, \big(\de\c\p_x+Q\c\p_x\big)\,,\label{Xx}
\ee 
where $Q$ does not contain any $x$ or $\p_x$. Plugging \eqref{Xx} into \eqref{eq}, one arrives at
\begin{align}
-\,2\,\p_x\c\p_x+2\,(\e\c\p_x)(\de\c\p_x)&=(\es+i\,)\,\ds\,(\de\c\p_x)+\le[\,(\es+i\,)\,\ds - (\e^2+1)(\de\c\p_x)\,\ri](Q\c\p_x)\,.
\end{align}
Since both sides of the latter are quadratic in $\p_x$, the symmetric part of the coefficients of $(\p_x^\a\,\p_x^\b)$ on the two sides should be equal, that is
\begin{align}
-\,4\,g^{\a\b}+2\,\big(\e^\a\,\de^\b+\e^\b\,\de^\a\big)&=\big(\es+i\,\big)\big(\g^\a\,\de^\b+\g^\b\,\de^\a\big)+\big(\es+i\,\big)\big(\g^\a\,Q^\b+\g^\b\,Q^\a\big)\label{ss}\\&-\,(\e^2+1)\,\big[(\de^\b\,Q^\a)+(\de^\a\,Q^\b)\big]-\,(\e^2+1)\,\big(Q^\a\,\de^\b+Q^\b\,\de^\a\big)\,.\nonumber
\end{align}
In this relation, if one equals the coefficients of the partial derivatives on the two sides, one arrives
at
\be 
Q^\a=\frac{(\es+i\,)\,\g^\a-2\,\e^\a}{\e^2+1}\,,\label{Q}
\ee 
while the remaining parts of \eqref{ss} leads to  
\be 
-\,4\,g^{\a\b}=\big(\es+i\,\big)\big(\g^\a\,Q^\b+\g^\b\,Q^\a\big)-\,(\e^2+1)\,\big[(\de^\b\,Q^\a)+(\de^\a\,Q^\b)\big]\,.\label{iden}
\ee
In conclusion, by plugging \eqref{Q} into \eqref{Xx} one can claim that the operator $X$ has been found, provided the identity \eqref{iden} holds. However, one can simply check that the identity \eqref{iden} does not satisfy if one uses \eqref{Q} in \eqref{iden}. This shows that the operator $X$ could not be found.

\section{Chiral multiplet $(\,{\scriptstyle 0}\,,\, {\scriptstyle 1/2}\,)$}\label{Wess-Zu}
Let us present here supersymmetry transformations for the Wess-Zumino model in which the bosonic field is a complex scalar field $\phi(x)$ and the ferminic one is a Weyl spinor $\psi_L(x)$ (left-handed one) or a Majorana spinor $\psi(x)$.

One can introduce left-handed and right-handed spinors $\psi_L(x)$, $\psi_R(x)$ satisfying respectively by
\be
\psi_L= \big(\tfrac{\,1\,+\,\g^5\,}{2}\big)\,\psi=L\,\psi\,,\qquad \qquad
\psi_R= \big(\tfrac{\,1\,-\,\g^5\,}{2}\big)\,\psi=R\,\psi\,.
\ee
Then, considering a left-handed spinor $\psi_L(x)$, the supersymmetry action is given by (for more details see \cite{Derendinger:1990tj})
\be 
S=\int\,d^4x~\le(-\,\phi^\dagger\,\Box\,\phi~+~\overline{\psi}_L\,i\,\ds\,\psi_L\ri)\,,
\ee 
which is invariant under the following supersymmetry transformations
\begin{align}
\d \,\phi &={\scriptstyle\sqrt{2}}~\overline{\epsilon_R}~{\psi}_L\,, \qquad\qquad\qquad
\d \,\phi^\dagger={\scriptstyle\sqrt{2}}~\overline{\psi_L}~{\epsilon_R}\,, \\
\d \,\psi_L &=-\,i\,{\scriptstyle\sqrt{2}}~ \ds~\phi~\epsilon_R\,,\qquad\qquad
\d \,\overline{\psi_L} =i\,{\scriptstyle\sqrt{2}}~\overline{\epsilon_R} ~\ds~\phi^\dagger\,.
\end{align}

Taking into account a Majorana spinor, the supersymmetry action is given by
\be 
S=\int\,d^4x~\le(-\,\phi^\dagger\,\Box\,\phi~+~\tfrac{1}{2}\,\overline{\psi}\,i\,\ds\,\psi\ri)\,,
\ee 
which is invariant under the supersymmetry transformations
\begin{align}
\d \,\phi&={\scriptstyle\sqrt{2}}~\bar\epsilon~\big(\tfrac{\,1\,+\,\g^5\,}{2}\big)~{\psi}\,,\qquad\qquad\qquad~~\,\qquad\qquad\qquad\d \,\phi^\dagger={\scriptstyle\sqrt{2}}~\overline{\psi}~\big(\tfrac{\,1\,-\,\g^5\,}{2}\big)~{\epsilon}\,, \\
\d \,\psi&=-\, \tfrac{i}{\sqrt{2}}\le[\,\ds\,(1-\g^5)\,\phi+\ds\,(1+\g^5)\,\phi^\dagger\,\ri]\epsilon\,,\qquad\qquad
\d \,\overline{\psi}= \tfrac{i}{\sqrt{2}}\,\bar\epsilon\le[\,\ds\,(1+\g^5)\,\phi+\ds\,(1-\g^5)\,\phi^\dagger\,\ri]\,.
\end{align}
If one defines the complex scalar field as
\be 
\phi=\tfrac{1}{\sqrt{2}}\,(A\,-\,i\,B)\,,
\ee 
where $A$ and $B$ are two real scalar fields, one can rewrite the supersymmetry action as
\be 
S=\frac{1}{2}\,\int\,d^4x~\le(-~A\,\Box\,A~-~\,B\,\Box\,B~+~\,\overline{\psi}\,i\,\ds\,\psi\ri)\,,
\ee 
which would be invariant under the following supersymmetry transformations
\begin{align}
\d \,A&=\bar\epsilon~{\psi}\,,\qquad\qquad\qquad~~\,\qquad\d \,\psi=-\,\big(\,i\,\ds\,A+\g^5\,\ds\,B\,\big)\,\epsilon\,, \\
\d \,B&=i\,\bar\epsilon\,\g^5\,\psi\,,\qquad\qquad\qquad~~~~
\d \,\overline{\psi}= \,\bar\epsilon\,\,\big(\,i\,\ds\,A-\g^5\,\ds\,B\,\big)\,.
\end{align}

\section{Inverse operators}\label{I op}

To make a relationship between the unconstrained and constrained supersymmetry transformations in section \ref{sec. rel}, we may need to know the inverse of operators in \eqref{P_phi text} and \eqref{P say}. Therefore, referring to \cite{Najafizadeh:2017acd} (appendix B), let us clarify and find the inverse operators in the following. We will follow this appendix in $d$-dimensional space-time to include a more general form. 

For {\it bosonic} fields, introducing the bosonic operators
\be
\mathbf{P}_\Phi:= \sum_{k=0}^{\infty}~\omega^{\,2k}~ \tfrac{1}{~4^{k} ~ k!~ (N+\frac{d}{2}-1)_k~}\,, \qquad \qquad
\mathbf{Q}_\Phi:= \sum_{k=0}^{\infty}~\omega^{\,2k}~ \tfrac{1}{~4^{k} ~ k!~ (N+\frac{d}{2}+3)_k~}\,,\label{op}
\ee 
where $N:={\w\c\dw}$ and $(a)_k$ is the rising Pochhammer symbol \eqref{Pochhammer}, one can show (see explanations of the appendix B in \cite{Najafizadeh:2017acd})
\be 
\mathbf{Q}_\Phi\,(\dww)^2=(\dww-1\,)^2\,\mathbf{P}_\Phi\,.\label{ident}
\ee 
Then, using the latter, and by the following field redefinition
\be 
\widetilde{\Phi} (x,\omega):=\mathbf{P}_\Phi~\Phi (x,\omega)\,,\label{def}
\ee 
one can convert the double trace-like constraint to the double trace one as the following 
\be 
(\dww-1\,)^2\,\widetilde{\Phi}=0\quad\quad \Longrightarrow\quad\quad
(\dww-1\,)^2\,\mathbf{P}_\Phi~\Phi=0 \quad\quad\Longrightarrow\quad\quad \mathbf{Q}_\Phi\,(\dww)^2\,\Phi=0\quad\quad\Longrightarrow\quad\quad (\dww)^2\,\Phi=0\,.
\ee 
Now the question is what happens if one wants to apply a reverse way in \eqref{def} and rewrite $\Phi$ in terms of $\widetilde\Phi$? In other words, one can define
\be 
{\Phi} (x,\omega):=\mathbf{P}^{-1}_\Phi~\widetilde\Phi (x,\omega)\,,
\ee 
and use the identity (which can be simply obtained from \eqref{ident}) 
\be 
(\dww)^2\,\,\mathbf{P}_\Phi^{-1}=\mathbf{Q}_\Phi^{-1}\,(\dww-1\,)^2\,,\label{revers}
\ee 
to rearrange the double trace constraint to the double trace-like one, using the latter identity, in the following form
\be
(\dww)^2\,\Phi=0\quad\quad \Longrightarrow\quad\quad
(\dww)^2\,\mathbf{P}_\Phi^{-1}\,\widetilde\Phi=0\quad\quad \Longrightarrow\quad\quad
\mathbf{Q}_\Phi^{-1}\,(\dww-1\,)^2\,\widetilde\Phi=0\quad\quad \Longrightarrow\quad\quad
(\dww-1\,)^2\,\widetilde\Phi=0\,.
\ee 
In this case, we would like to know inverse operators which will satisfy the relation \eqref{revers}. It is convenient to find that those inverse bosonic operators satisfying \eqref{revers} are
\be
\mathbf{P}_\Phi^{-1}:= \sum_{k=0}^{\infty}~\omega^{\,2k}~ \tfrac{(-1)^k}{~4^{k} ~ k!~ (N+\frac{d}{2}+k-2)_k~}\,, \qquad \qquad
\mathbf{Q}_\Phi^{-1}:= \sum_{k=0}^{\infty}~\omega^{\,2k}~ \tfrac{(-1)^k}{~4^{k} ~ k!~ (N+\frac{d}{2}+k+2)_k~}\,.
\ee 
We note that these obtained reverse operators can be also acquired directly from \eqref{op}.

\vspace{.5cm}

For {\it fermionic} fields, as we mentioned, there are two possibilities to present the action principle and so on. Therefore, one can introduce 
\be 
\mathbf{P}_\Psi^{\pm} :=\sum_{k=0}^{\infty}~\le[(\ws)^{2k} \pm 2k(\ws)^{2k-1}\ri]
\tfrac{1}{~4^{k}~k!~(N+\frac{d}{2}-1)_k~}\,, \qquad 
\mathbf{Q}_\Psi^{\pm} :=\sum_{k=0}^{\infty}~\le[(\ws)^{2k} \mp 2k(\ws)^{2k-1}\ri]
\tfrac{1}{~4^{k}~k!~(N+\frac{d}{2}+2)_k~}\,,
\ee 
where the upper and lower signs are related to each possibility we choose. These two set of operators satisfy the following two possible identities
\be 
\mathbf{Q}_\Psi^+\,(\dws)^3=(\dws-1\,)(\dww-1)\,\mathbf{P}_\Psi^+\,,\qquad\qquad
\mathbf{Q}_\Psi^-\,(\dws)^3=(\dws+1\,)(\dww-1)\,\mathbf{P}_\Psi^-\,.
\ee 
Therefore, there are two possible ways to apply a field redefinition. Indeed, by applying the field redefinitions
\be 
\widetilde{\Psi} (x,\omega):=\mathbf{P}_\Psi^+~\Psi (x,\omega)\,,\quad\quad\quad\quad\hbox{or}\quad\quad\quad\quad
\widetilde{\Psi} (x,\omega):=\mathbf{P}_\Psi^-~\Psi (x,\omega)\,,\label{def 2}
\ee 
one can convert the two possible triple gamma-trace-like conditions to the triple gamma-trace condition
\be 
(\dws\mp 1\,)(\dww-1)\,\widetilde{\Psi}=0\quad\quad \Longrightarrow\quad
(\dws\mp1\,)(\dww-1)\,\mathbf{P}_\Psi^{\pm}~\Psi=0\quad\quad \Longrightarrow\quad
\mathbf{Q}_\Psi^{\pm}\,(\dws)^3\,\Psi=0\quad\quad \Longrightarrow\quad
(\dws)^3\,\Psi=0\,.\nonumber
\ee 
If we follow a similar manner as the bosonic case, we can conveniently find the inverse fermionic operators, which are
\bea 
(\mathbf{P}_\Psi^{\pm})^{-1} &:=&\sum_{k=0}^{\infty}
\tfrac{(-1)^k}{~4^{k}~k!~(N+\frac{d}{2}-k-1)_k~}\le[(\ws)^{2k} \pm 2k(\ws)^{2k-1}\ri]\,,\label{p} \\ 
(\mathbf{Q}_\Psi^{\pm})^{-1} &:=&\sum_{k=0}^{\infty}
\tfrac{(-1)^k}{~4^{k}~k!~(N+\frac{d}{2}-k+2)_k~}\le[(\ws)^{2k} \mp 2k(\ws)^{2k-1}\ri]\,.
\eea 
We note that in this paper and \cite{Najafizadeh:2018cpu} we have chose the upper sign of the fermionic operators, while we have considered the lower sign in \cite{Najafizadeh:2015uxa,Bekaert:2017xin} in order to rearrange trace conditions on the fermionic CSP or higher spin fields. In $d=4$ dimensions, by choosing the upper sign and keeping the terms up to $\mathcal{O}(\w^3)$, one can write \eqref{p} as
\be 
(\mathbf{P}_\Psi^{+})^{-1}=1\,-\,\ws\,\tfrac{1}{2(N+1)}\,-\,\w^2\,\tfrac{1}{4(N+2)}\,+\,\mathcal{O}(\w^3)\,,\label{ps-1}
\ee 
which can be used in section \ref{sec. rel}. Note that the upper sign is left implicit in \eqref{P say}.

\section{Useful relations}\label{usef}

The ``Majorana flip relations'' or the so-called ``Dirac flip relations'' are given by
\begin{align}
\bar\epsilon_2\,(\g^{\m_1}\,\g^{\m_2}\,\cdots\,\g^{\m_p})\,\epsilon_1&=(-1)^p~\bar\epsilon_1\,(\g^{\m_1}\,\g^{\m_2}\,\cdots\,\g^{\m_p})\,\epsilon_2 \,, \label{maj flip}
\end{align}
where $\epsilon_1$ and $\epsilon_2$ can be either the Majorana spinors or Dirac spinors depending on the problem we are dealing with (see \cite{Freedman:2012zz}, page 49, for more details).

\vspace{.3cm}

In order to illustrate that the SUSY HS/CSP actions are invariant under supersymmetry transformations, the following obtained relations are useful:
\bea
\hspace{-.75cm}\wdx\dws(\ws\ds\dws)&=&\wdx\ws\ds\dww-2\,\ws\wdx\dwdx\dws+2\,\wdx \,h\, \ds\dws\label{01}
\\[5pt]
\hspace{-.75cm}(\ws\ds\dws)(\ws\ds\dws)&=&-\,\w^2\,\Box\,\dww+2\,\ws\Box\, h\, \dws+2\,\w^2\ds \dwdx\dws+2\,\ws\wdx\ds\,\dww-4\,\ws\wdx\dwdx\dws
\\[5pt] 
\hspace{-.75cm}\tfrac{1}{2}\,\ws\wdx\dww(\ws\ds\dws)&=&-\,\ws\wdx\ds\dww+2\,\ws\wdx\dwdx\dws+\tfrac{1}{2}\,\w^2\wdx\ds\dws^3
\\[5pt] 
\hspace{-.75cm}\tfrac{1}{2}\,\w^2\dwdx\dws(\ws\ds\dws)&=&\tfrac{1}{2}\,\w^2\Box\dww-\w^2\ds\dwdx\dws+\w^2\ds(h+1)\dwdx\dws
\\[5pt] 
\hspace{-.75cm}-\,\tfrac{1}{4}\,\w^2\ds\dww(\ws\ds\dws)&=&\tfrac{1}{2}\,\w^2\Box\dww-\w^2\ds\dwdx\dws-\tfrac{1}{2}\,\w^2\wdx\ds\dws^3
\\[5pt] 
\hspace{-.75cm}(\ws\ds\dws)(\ws\dws)&=&\w^2\ds\dww-2\,\ws\wdx\dww+2\,\ws\ds h\dws
\\[5pt]
\hspace{-.75cm}(\ws\ds\dws)(\w^2\dww)&=&-2\,\w^2\ds\dww+4\,\ws\wdx\dww
\\[5pt] 
\hspace{-.75cm}\tfrac{1}{2}\,\w^2\dwdx\dws(\ws\dws)&=&-\,\tfrac{1}{2}\,\w^2\ds\dww+\w^2(h+1)\dwdx\dws
\\[5pt]
\hspace{-.75cm}\tfrac{1}{2}\,\w^2\dwdx\dws(\w^2\dww)&=&\w^2\ds\dww+\w^2 \wdx\dws^3
\\[5pt] 
\hspace{-.75cm}-\,\tfrac{1}{4}\,\w^2\ds\dww (\ws\dws)&=&-\,\tfrac{1}{2}\,\w^2\wdx\dws^3-\,\tfrac{1}{2}\,\w^2\ds\dww \label{10}
\eea
where $h:=\w\c\dw+\frac{d}{2}$\,, such that $d$ is space-time dimension. We obtained these relations using (anti-)commutation relations presented in appendix of \cite{Najafizadeh:2017acd} and have dropped terms of order $\mathcal{O}(\w^3)$ and $\mathcal{O}(\dw^4)$ which vanish by constraints $\overline{\psi}(x,\dw)\,\ws^3=0$ and $(\dww)^2\,\phi(x,\w)=0$ in the SUSY actions.

\vspace{.3cm}

In order to close the SUSY algebra, one can use the identity
\be 
\g^\m\g^\n\g^\r=\e^{\m\n}\g^\r+\g^\m \e^{\n\r}- \e^{\m\r}\g^\n - i\,\epsilon^{\a\m\n\r}\,\g_\a\,\g^5\,,
\ee 
leading to
\be 
\ws \ds \dws = (\w\c\p_x)\,\dws+\ws(\dw\c\p_x) - N\,\ds -i\,\epsilon^{\a\m\n\r}\,\w_\m \p_\n {\dw}_\r (\g_\a\,\g^5)\,,\label{123}
\ee 
where $N:=\w\c\dw$. Moreover, defining 
\bea 
\mathcal{A}_{21}&:=&(1-\g^5\,)\,\epsilon_2\,\bar\epsilon_1\,(1+\g^5\,)\,,\\
\mathcal{B}_{21}&:=&(1+\g^5\,)\,\epsilon_2\,\bar\epsilon_1\,(1-\g^5\,)\,,
\eea 
one can show
\bea 
\mathcal{A}_{21} -\mathcal{A}_{12} &=&-\,\bar\epsilon_1\,\g_\m\,\epsilon_2\,\g^\m\,(1+\g^5\,)\,,\\
\mathcal{B}_{21} -\mathcal{B}_{12} &=&-\,\bar\epsilon_1\,\g_\m\,\epsilon_2\,\g^\m\,(1-\g^5\,)\,,
\eea 
where we have applied the identity
\be 
\epsilon_2\,\bar\epsilon_1-\epsilon_1\,\bar\epsilon_2-\g^5\,(\epsilon_2\,\bar\epsilon_1-\epsilon_1\,\bar\epsilon_2)\,\g^5=-\,\bar\epsilon_1\,\g_\m\,\epsilon_2\,\g^\m\,.
\ee 

\vspace{.3cm}

The \textit{rising Pochhammer symbol} $(a)_n$ is defined as
\be
(a)_n := 
a \, (a+1)(a+2) \cdots (a+n-1) = \frac{\Gamma(a+n)}{\Gamma(a)} \,, \quad\quad n\in\mathbb N ~~\mbox{and}~~ a\in\mathbb R\,.  \label{Pochhammer}
\ee

\vspace{.3cm}

It is convenient to show that operators $\mathbf{P}_\Phi$ and $\mathbf{P}_\Psi$, defined in \eqref{P_phi text} and \eqref{P say}, are related to each other through 
\be
\mathbf{P}_{\Psi} = \mathbf{P}_\Phi \le( 1 +\,\ws ~\tfrac{1}{2(N+1)} \ri)\,. \label{psiphi}
\ee

The quantities $\dw^{\,\a}$\,, $\dw^{\,2}$ and $\w^\a$ act on the bosonic operator $\mathbf{P}_\Phi$ \eqref{P_phi text} as (for details, see the appendices in \cite{Najafizadeh:2017acd}\,{\color{blue}\footnote{We note, in this paper, we work in 4 dimensions and the metric signature is considered to be mostly minus, while in \cite{Najafizadeh:2017acd}, we have chosen the metric signature mostly plus and presented results in $d$ dimensions.}})
\begin{align}
~~\dw^{\,\a}~\mathbf{P}_\Phi&=\mathbf{P}_\Phi\le[\, \dw^{\,\a}\,-\,\w^2~\tfrac{1}{4(N+1)(N+2)}~\dw^{\,\a}\,+\,\w^\a~\tfrac{1}{2(N+1)}\,\ri]\,,  \label{11}
\\[10pt]
\dw^{\,2}~\mathbf{P}_\Phi&=\mathbf{P}_\Phi\le[\,\dww\,-\,\w^2~\tfrac{1}{2(N+1)(N+3)}~\dww
\,+\,\tfrac{N+2}{\,\,N+1\,\,} \,-\,\w^2~\tfrac{1}{4(N+2)(N+1)^2}+~  \mathcal{O}(\w^4)\, \ri]\,, \label{22}
\\[10pt]
\w^a~\mathbf{P}_\Phi&=\mathbf{P}_\Phi\le[\,\w^\a\,+\,\w^2\, \w^\a~\tfrac{1}{4(N+1)(N+2)}~+~  \mathcal{O}(\w^4)\, \ri]\,, \label{33}
\end{align}
where the terms containing $\mathcal{O}(\w^4)$ in two last relations will be eliminated at the level of the action, due to the double-traceless condition on the gauge field ${\Phi}(x,\dw)\,(\w^2)^{\,2}\,=0$\,.



\vspace{.5cm}
\noindent\textbf{\centerline{References:}} 

\vspace{-.5cm}
\begin{thebibliography}{99}

	
\bibitem{Wigner}
E.P.~Wigner, ``On Unitary Representations of the Inhomogeneous Lorentz Group'', \href{http://inspirehep.net/record/26312?ln=en}{\emph{Ann. Math.} {\bf 40} (1939) 149}.

\bibitem{Unitary}
X.~Bekaert and N.~Boulanger, ``The Unitary representations of the Poincare group in any spacetime
dimension'', \href{http://inspirehep.net/record/732636?ln=en}{hep-th/0611263}.

\bibitem{Brink:2002zx}
L.~Brink, A.~M.~Khan, P.~Ramond and X.~z.~Xiong,
``Continuous spin representations of the Poincare and super-Poincare groups'',
\emph{J. Math. Phys.}  {\bf 43}, (2002) 6279
[\href{http://inspirehep.net/record/586869?ln=en}{hep-th/0205145}].	
	
\bibitem{Schuster:2014hca} 
P.~Schuster and N.~Toro,
``Continuous-spin particle field theory with helicity correspondence,''
Phys.\ Rev.\ D {\bf 91}, 025023 (2015)
[\href{http://inspirehep.net/record/1288515?ln=en}{arXiv:1404.0675}].


\bibitem{Najafizadeh:2015uxa} 
X.~Bekaert, M.~Najafizadeh and M.~R.~Setare,
``A gauge field theory of fermionic Continuous-Spin Particles,''
Phys.\ Lett.\ B {\bf 760}, 320 (2016)
[\href{http://inspirehep.net/record/1374250?ln=en}{arXiv:1506.00973}].



\bibitem{Metsaev:2016lhs} 
R.~R.~Metsaev,
``Continuous spin gauge field in (A)dS space,''
Phys.\ Lett.\ B {\bf 767}, 458 (2017)
[\href{http://inspirehep.net/record/1489236?ln=en}{arXiv:1610.00657}].


\bibitem{Metsaev:2017ytk} 
R.~R.~Metsaev,
``Fermionic continuous spin gauge field in (A)dS space,''
Phys.\ Lett.\ B {\bf 773}, 135 (2017)
[\href{http://inspirehep.net/record/1518169?ln=en}{arXiv:1703.05780}].


\bibitem{Bekaert:2017xin} 
X.~Bekaert, J.~Mourad and M.~Najafizadeh,
``Continuous-spin field propagator and interaction with matter,''
JHEP {\bf 1711}, 113 (2017)
[\href{http://inspirehep.net/record/1630946?ln=en}{arXiv:1710.05788}].

\bibitem{Najafizadeh:2017acd} 
M.~Najafizadeh,
``Modified Wigner equations and continuous spin gauge field,''
Phys.\ Rev.\ D {\bf 97}, no. 6, 065009 (2018)
[\href{http://inspirehep.net/record/1614183?ln=en}{arXiv:1708.00827}].


\bibitem{Bekaert:2005in} 
X.~Bekaert and J.~Mourad,
``The Continuous spin limit of higher spin field equations,''
JHEP {\bf 0601}, 115 (2006)
[\href{http://inspirehep.net/record/692077?ln=en}{hep-th/0509092}].


\bibitem{Segal:2001qq} 
A.~Y.~Segal,
``A Generating formulation for free higher spin massless fields,''
{\href{http://inspirehep.net/record/553784?ln=en}{hep-th/0103028}}.


\bibitem{Najafizadeh:2018cpu}
M.~Najafizadeh,
``Local action for fermionic unconstrained higher spin gauge fields in AdS and dS spacetimes,''
Phys.\ Rev.\ D {\bf 98}, no. 12, 125012 (2018)
[\href{http://inspirehep.net/record/1680667?ln=en}{arXiv:1807.01124}].


\bibitem{Fronsdal:1978rb}C.~Fronsdal,
``Massless Fields with Integer Spin,''
\href{http://inspirehep.net/record/130533?ln=en}{Phys.\ Rev.\ D {\bf 18}, 3624 (1978)}.

\bibitem{Fang:1978wz} 
J.~Fang and C.~Fronsdal,
``Massless Fields with Half Integral Spin,''
\href{http://inspirehep.net/record/131028?ln=en}{Phys.\ Rev.\ D {\bf 18}, 3630 (1978)}.




\bibitem{Zinoviev:2017rnj} 
Y.~M.~Zinoviev,
``Infinite spin fields in d = 3 and beyond,''
Universe {\bf 3}, no. 3, 63 (2017)
[\href{http://inspirehep.net/record/1613343?ln=en}{arXiv:1707.08832}].

\bibitem{Buchbinder:2019kuh} 
I.~L.~Buchbinder, M.~V.~Khabarov, T.~V.~Snegirev and Y.~M.~Zinoviev,
``Lagrangian formulation for the infinite spin $N$=1 supermultiplets in $d$=4,''
Nucl.\ Phys.\ B {\bf 946}, 114717 (2019)
[\href{http://inspirehep.net/record/1729257?ln=en}{arXiv:1904.05580}].


\bibitem{Buchbinder:2019iwi} 
I.~L.~Buchbinder, S.~Fedoruk and A.~P.~Isaev,
``Twistorial and space-time descriptions of massless infinite spin (super)particles and fields,''
Nucl.\ Phys.\ B {\bf 945}, 114660 (2019)
[\href{http://inspirehep.net/record/1725750?ln=en}{arXiv:1903.07947}].

\bibitem{Buchbinder:2019esz} 
I.~L.~Buchbinder, S.~J.~Gates and K.~Koutrolikos,
``Superfield continuous spin equations of motion,''
Phys.\ Lett.\ B {\bf 793}, 445 (2019)
[\href{http://inspirehep.net/record/1725940?ln=en}{arXiv:1903.08631}].

\bibitem{Buchbinder:2019sie} 
I.~L.~Buchbinder, S.~Fedoruk and A.~P.~Isaev,
``Massless infinite spin (super)particles and fields,''
\href{http://inspirehep.net/record/1762678?ln=en}{arXiv:1911.00362}.


\bibitem{Khan:2004nj} 
A.~M.~Khan and P.~Ramond,
``Continuous spin representations from group contraction,''
J.\ Math.\ Phys.\  {\bf 46}, 053515 (2005)
Erratum: [J.\ Math.\ Phys.\  {\bf 46}, 079901 (2005)]
[\href{http://inspirehep.net/record/661605?ln=en}{hep-th/0410107}].

\bibitem{Edgren:2005gq} 
L.~Edgren, R.~Marnelius and P.~Salomonson,
``Infinite spin particles,''
JHEP {\bf 0505}, 002 (2005)
[\href{http://inspirehep.net/record/678634?ln=en}{hep-th/0503136}].


%

\bibitem{Edgren:2006un} 
L.~Edgren and R.~Marnelius,
``Covariant quantization of infinite spin particle models, and higher order gauge theories,''
JHEP {\bf 0605}, 018 (2006)
[\href{http://inspirehep.net/record/710175?ln=en}{hep-th/0602088}].


\bibitem{Mourad:2006xk} 
J.~Mourad,
``Continuous spin particles from a tensionless string theory,''
\href{http://inspirehep.net/record/736600?ln=en}{AIP Conf.\ Proc.\  {\bf 861}, no. 1, 436 (2006)}.

\bibitem{Schuster:2013pxj} 
P.~Schuster and N.~Toro,
``On the Theory of Continuous-Spin Particles: Wavefunctions and Soft-Factor Scattering Amplitudes,''
JHEP {\bf 1309}, 104 (2013)
[\href{http://inspirehep.net/record/1217871?ln=en}{arXiv:1302.1198}].


\bibitem{Schuster:2013vpr} 
P.~Schuster and N.~Toro,
``On the Theory of Continuous-Spin Particles: Helicity Correspondence in Radiation and Forces,''
JHEP {\bf 1309}, 105 (2013)
[\href{http://inspirehep.net/record/1218027?ln=en}{arXiv:1302.1577}].


\bibitem{Bengtsson:2013vra} 
A.~K.~H.~Bengtsson,
``BRST Theory for Continuous Spin,''
JHEP {\bf 1310}, 108 (2013)
[\href{http://inspirehep.net/record/1224023?ln=en}{arXiv:1303.3799}].


\bibitem{Schuster:2014xja} 
P.~Schuster and N.~Toro,
``A new class of particle in 2 + 1 dimensions,''
Phys.\ Lett.\ B {\bf 743}, 224 (2015)
[\href{http://inspirehep.net/record/1288713?ln=en}{arXiv:1404.1076}].



\bibitem{Rivelles:2014fsa} 
V.~O.~Rivelles,
``Gauge Theory Formulations for Continuous and Higher Spin Fields,''
Phys.\ Rev.\ D {\bf 91}, no. 12, 125035 (2015)
[\href{http://inspirehep.net/record/1311237?ln=en}{arXiv:1408.3576}].


\bibitem{Schroer:2015rct} 
B.~Schroer,
``Wigner’s infinite spin representations and inert matter,''
Eur.\ Phys.\ J.\ C {\bf 77}, no. 6, 362 (2017)
[\href{http://inspirehep.net/record/1478648?ln=en}{arXiv:1601.02477}].

\bibitem{Rivelles:2016rwo} 
V.~O.~Rivelles,
``Remarks on a Gauge Theory for Continuous Spin Particles,''
Eur.\ Phys.\ J.\ C {\bf 77}, no. 7, 433 (2017)
[\href{http://inspirehep.net/record/1473857?ln=en}{arXiv:1607.01316}].





\bibitem{Bekaert:2017khg} 
X.~Bekaert and E.~D.~Skvortsov,
``Elementary particles with continuous spin,''
Int.\ J.\ Mod.\ Phys.\ A {\bf 32}, no. 23n24, 1730019 (2017)
[\href{http://inspirehep.net/record/1614342?ln=en}{arXiv:1708.01030}].


\bibitem{Rehren:2017xzn} 
K.~H.~Rehren,
``Pauli-Lubanski limit and stress-energy tensor for infinite-spin fields,''
JHEP {\bf 1711}, 130 (2017)
[\href{http://inspirehep.net/record/1623580?ln=en}{arXiv:1709.04858}].



\bibitem{Metsaev:2017cuz} 
R.~R.~Metsaev,
``Cubic interaction vertices for continuous-spin fields and arbitrary spin massive fields,''
JHEP {\bf 1711}, 197 (2017)
[\href{http://inspirehep.net/record/1625344?ln=en}{arXiv:1709.08596}].


	
	

\bibitem{Khabarov:2017lth} 
M.~V.~Khabarov and Y.~M.~Zinoviev,
``Infinite (continuous) spin fields in the frame-like formalism,''
Nucl.\ Phys.\ B {\bf 928}, 182 (2018)
[\href{http://inspirehep.net/record/1637612?ln=en}{arXiv:1711.08223}].

\bibitem{Gracia-Bondia:2017fai} 
J.~M.~Gracia-Bondia, F.~Lizzi, J.~C.~Varilly and P.~Vitale,
``The Kirillov picture for the Wigner particle,''
J.\ Phys.\ A {\bf 51}, no. 25, 255203 (2018)
[\href{http://inspirehep.net/record/1639050?ln=en}{arXiv:1711.09608}].


\bibitem{Metsaev:2017myp} 
R.~R.~Metsaev,
``Continuous-spin mixed-symmetry fields in AdS(5),''
J.\ Phys.\ A {\bf 51}, no. 21, 215401 (2018)
[\href{http://inspirehep.net/record/1639482?ln=en}{arXiv:1711.11007}].

\bibitem{Alkalaev:2017hvj} 
K.~B.~Alkalaev and M.~A.~Grigoriev,
``Continuous spin fields of mixed-symmetry type,''
JHEP {\bf 1803}, 030 (2018)
[\href{http://inspirehep.net/record/1641316?ln=en}{arXiv:1712.02317}].

\bibitem{Metsaev:2018lth} 
R.~R.~Metsaev,
``BRST-BV approach to continuous-spin field,''
Phys.\ Lett.\ B {\bf 781}, 568 (2018)
[\href{http://inspirehep.net/record/1663599?ln=en}{arXiv:1803.08421}].


\bibitem{Buchbinder:2018soq} 
I.~L.~Buchbinder, S.~Fedoruk, A.~P.~Isaev and A.~Rusnak,
``Model of massless relativistic particle with continuous spin and its twistorial description,''
JHEP {\bf 1807}, 031 (2018)
[\href{http://inspirehep.net/record/1674729?ln=en}{arXiv:1805.09706}].


\bibitem{Buchbinder:2018yoo} 
I.~L.~Buchbinder, V.~A.~Krykhtin and H.~Takata,
``BRST approach to Lagrangian construction for bosonic continuous spin field,''
Phys.\ Lett.\ B {\bf 785}, 315 (2018)
[\href{http://inspirehep.net/record/1676474?ln=en}{arXiv:1806.01640}].


\bibitem{Rivelles:2018tpt} 
V.~O.~Rivelles,
``A Gauge Field Theory for Continuous Spin Tachyons,''
\href{http://inspirehep.net/record/1681143?ln=en}{arXiv:1807.01812}.



\bibitem{Alkalaev:2018bqe} 
K.~Alkalaev, A.~Chekmenev and M.~Grigoriev,
``Unified formulation for helicity and continuous spin fermionic fields,''
JHEP {\bf 1811}, 050 (2018)
[\href{http://inspirehep.net/record/1691814?ln=en}{arXiv:1808.09385}].


\bibitem{Gracia-Bondia:2018hrq} 
J.~M.~Gracia-Bondía and J.~C.~Várilly,
``On the kinematics of the last Wigner particle,''
Springer Proc.\ Phys.\  {\bf 229}, 225 (2019)
[\href{http://inspirehep.net/record/1692625?ln=en}{arXiv:1809.00387}].


\bibitem{Metsaev:2018moa} 
R.~R.~Metsaev,
``Cubic interaction vertices for massive/massless continuous-spin fields and arbitrary spin fields,''
JHEP {\bf 1812}, 055 (2018)
[\href{http://inspirehep.net/record/1695270?ln=en}{arXiv:1809.09075}].

\bibitem{Metsaev:2019opn} 
R.~R.~Metsaev,
``Light-cone continuous-spin field in AdS space,''
Phys.\ Lett.\ B {\bf 793}, 134 (2019)
[\href{http://inspirehep.net/record/1726544?ln=en}{arXiv:1903.10495}].


\bibitem{Burdik:2019tzg} 
C.~Burdík, V.~K.~Pandey and A.~Reshetnyak,
``BRST-BFV and BRST-BV Lagrangians for Bosonic Fields with Continuous Spin on $R^{1,d-1}$,''
\href{http://inspirehep.net/record/1738693?ln=en}{arXiv:1906.02585}.


\bibitem{Bekaert:2009pt} 
X.~Bekaert, M.~Rausch de Traubenberg and M.~Valenzuela,
``An infinite supermultiplet of massive higher-spin fields,''
JHEP {\bf 0905}, 118 (2009)
[\href{https://inspirehep.net/record/818101}{arXiv:0904.2533}].



\bibitem{Zinoviev:2007js} 
Y.~M.~Zinoviev,
``Massive N=1 supermultiplets with arbitrary superspins,''
Nucl.\ Phys.\ B {\bf 785}, 98 (2007)
{[\href{http://inspirehep.net/record/748403?ln=en}{arXiv:0704.1535}]}.



\bibitem{Curtright:1979uz} 
T.~Curtright,
``Massless Field Supermultiplets With Arbitrary Spin,''
\href{http://inspirehep.net/record/140820?ln=en}{Phys.\ Lett.\  {\bf 85B}, 219 (1979)}.


\bibitem{Vasiliev:1980as} 
M.~A.~Vasiliev,
``'gauge' Form Of Description Of Massless Fields With Arbitrary Spin. (in Russian),''
Yad.\ Fiz.\  {\bf 32}, 855 (1980)
[\href{http://inspirehep.net/record/159952?ln=en}{Sov.\ J.\ Nucl.\ Phys.\  {\bf 32}, 439 (1980)}].


\bibitem{Kuzenko:1993jp} 
S.~M.~Kuzenko, A.~G.~Sibiryakov and V.~V.~Postnikov,
``Massless gauge superfields of higher half integer superspins,''
JETP Lett.\  {\bf 57}, 534 (1993)
[\href{http://inspirehep.net/record/364787?ln=en}{Pisma Zh.\ Eksp.\ Teor.\ Fiz.\  {\bf 57}, 521 (1993)}].



\bibitem{Kuzenko:1993jq} 
S.~M.~Kuzenko and A.~G.~Sibiryakov,
``Massless gauge superfields of higher integer superspins,''
JETP Lett.\  {\bf 57}, 539 (1993)
[\href{http://inspirehep.net/record/364788?ln=en}{Pisma Zh.\ Eksp.\ Teor.\ Fiz.\  {\bf 57}, 526 (1993)}].


\bibitem{Peskin:1995ev} 
M.~E.~Peskin and D.~V.~Schroeder,
``\href{http://inspirehep.net/record/407703?ln=en}{An Introduction to quantum field theory},'' Reading, USA: Addison-Wesley (1995).


\bibitem{Freedman:2012zz} 
D.~Z.~Freedman and A.~Van Proeyen,
``\href{http://inspirehep.net/record/1123253?ln=en}{Supergravity},'' Cambridge, UK: Cambridge Univ. Press (2012)

\bibitem{Francia:2002aa} 
D.~Francia and A.~Sagnotti,
``Free geometric equations for higher spins,''
Phys.\ Lett.\ B {\bf 543}, 303 (2002)
[\href{https://inspirehep.net/record/589618}{hep-th/0207002}].


\bibitem{Francia:2002pt} 
D.~Francia and A.~Sagnotti,
``On the geometry of higher spin gauge fields,''
Class.\ Quant.\ Grav.\  {\bf 20}, S473 (2003)
[Comment.\ Phys.\ Math.\ Soc.\ Sci.\ Fenn.\  {\bf 166}, 165 (2004)]
[PoS JHW {\bf 2003}, 005 (2003)]
[\href{https://inspirehep.net/record/604719}{hep-th/0212185}].

\bibitem{Francia:2007qt} 
D.~Francia, J.~Mourad and A.~Sagnotti,
``Current Exchanges and Unconstrained Higher Spins,''
Nucl.\ Phys.\ B {\bf 773}, 203 (2007)
[\href{https://inspirehep.net/record/742681}{hep-th/0701163}].


\bibitem{Buchbinder:2007ak} 
I.~L.~Buchbinder, A.~V.~Galajinsky and V.~A.~Krykhtin,
``Quartet unconstrained formulation for massless higher spin fields,''
Nucl.\ Phys.\ B {\bf 779}, 155 (2007)
[\href{https://inspirehep.net/record/744923}{hep-th/0702161}].


\bibitem{Buchbinder:2008ss} 
I.~L.~Buchbinder and A.~V.~Galajinsky,
``Quartet unconstrained formulation for massive higher spin fields,''
JHEP {\bf 0811}, 081 (2008)
[\href{https://inspirehep.net/record/799586}{arXiv:0810.2852}].


\bibitem{Francia:2013sca} 
D.~Francia, S.~L.~Lyakhovich and A.~A.~Sharapov,
``On the gauge symmetries of Maxwell-like higher-spin Lagrangians,''
Nucl.\ Phys.\ B {\bf 881}, 248 (2014)
[\href{https://inspirehep.net/record/1262838}{arXiv:1310.8589}].



\bibitem{Francia:2012rg}
D.~Francia,
``Generalised connections and higher-spin equations,''
Class.\ Quant.\ Grav.\  {\bf 29} (2012) 245003
[\href{https://inspirehep.net/record/1187654}{arXiv:1209.4885}].


	


\bibitem{Sorokin:2017irs} 
D.~Sorokin and M.~Tsulaia,
``Higher Spin Fields in Hyperspace. A Review,''
Universe {\bf 4}, no. 1, 7 (2018)
[\href{https://inspirehep.net/record/1632040}{arXiv:1710.08244}].


\bibitem{Bandos:1999qf} 
I.~A.~Bandos, J.~Lukierski and D.~P.~Sorokin,
``Superparticle models with tensorial central charges,''
Phys.\ Rev.\ D {\bf 61}, 045002 (2000)
[\href{https://inspirehep.net/record/498374}{hep-th/9904109}].


\bibitem{Bandos:2004nn} 
I.~Bandos, P.~Pasti, D.~Sorokin and M.~Tonin,
``Superfield theories in tensorial superspaces and the dynamics of higher spin fields,''
JHEP {\bf 0411}, 023 (2004)
[\href{https://inspirehep.net/record/654950}{hep-th/0407180}].


\bibitem{Florakis:2014kfa} 
I.~Florakis, D.~Sorokin and M.~Tsulaia,
``Higher Spins in Hyperspace,''
JHEP {\bf 1407}, 105 (2014)
[\href{https://inspirehep.net/record/1276428}{arXiv:1401.1645}].


\bibitem{Florakis:2014aaa} 
I.~Florakis, D.~Sorokin and M.~Tsulaia,
``Higher Spins in Hyper-Superspace,''
Nucl.\ Phys.\ B {\bf 890}, 279 (2014)
[\href{https://inspirehep.net/record/1312792}{arXiv:1408.6675}].


\bibitem{Metsaev:2019dqt} 
R.~R.~Metsaev,
``Cubic interaction vertices for N=1 arbitrary spin massless supermultiplets in flat space,''
JHEP {\bf 1908}, 130 (2019)
[\href{http://inspirehep.net/record/1736958?ln=en}{arXiv:1905.11357}].



\bibitem{Metsaev:2019aig} 
R.~R.~Metsaev,
``Cubic interactions for arbitrary spin $ \mathcal{N} $ -extended massless supermultiplets in 4d flat space,''
JHEP {\bf 1911}, 084 (2019)
[\href{http://inspirehep.net/record/1753649?ln=en}{arXiv:1909.05241}].

\bibitem{Derendinger:1990tj} 
J.~P.~Derendinger,
``\href{http://people.roma2.infn.it/~fucito/appunti/derendinger.pdf}{Lecture notes on globally supersymmetric theories in four and two dimensions},''
\href{http://inspirehep.net/record/300585?ln=en}{ETH-TH-90-21}.



	
	
	
	




	







	
\end{thebibliography}
 \end{document}